\documentclass[12pt]{article}

\usepackage{graphicx}
\usepackage{graphics}
\usepackage{amssymb}
\usepackage{amsmath}

\usepackage{epsf,amsfonts,hyperref}
\newcommand\encadremath[1]{\vbox{\hrule\hbox{\vrule\kern8pt 
\vbox{\kern8pt \hbox{$\displaystyle #1$}\kern8pt} 
\kern8pt\vrule}\hrule}}
\def\enca#1{\vbox{\hrule\hbox{
\vrule\kern8pt\vbox{\kern8pt \hbox{$\displaystyle #1$}
\kern8pt} \kern8pt\vrule}\hrule}}

\newcommand\figureframex[3]{
\begin{figure}[bth]
\hrule\hbox{\vrule\kern8pt 
\vbox{\kern8pt \vbox{
\begin{center}
{\mbox{\epsfxsize=#1.truecm\epsfbox{#2}}}
\end{center}
\caption{#3}
}\kern8pt} 
\kern8pt\vrule}\hrule
\end{figure}
}
\newcommand\figureframey[3]{
\begin{figure}[bth]
\hrule\hbox{\vrule\kern8pt 
\vbox{\kern8pt \vbox{
\begin{center}
{\mbox{\epsfysize=#1.truecm\epsfbox{#2}}}
\end{center}
\caption{#3}
}\kern8pt} 
\kern8pt\vrule}\hrule
\end{figure}
}

\makeatletter
\@addtoreset{equation}{section}
\makeatother
\newtheorem{theorem}{Theorem}[section]

\newtheorem{remark}{Remark}[section]
\newtheorem{proposition}{Proposition}[section]
\newtheorem{lemma}{Lemma}[section]
\newtheorem{corollary}{Corollary}[section]
\newtheorem{definition}{Definition}[section]
\def\br{\begin{remark}\rm\small}
\def\er{\end{remark}}
\def\bt{\begin{theorem}}
\def\et{\end{theorem}}
\def\bd{\begin{definition}}
\def\ed{\end{definition}}
\def\bp{\begin{proposition}}
\def\ep{\end{proposition}}
\def\bl{\begin{lemma}}
\def\el{\end{lemma}}
\def\bc{\begin{corollary}}
\def\ec{\end{corollary}}
\def\beaq{\begin{eqnarray}}
\def\eeaq{\end{eqnarray}}

\newcommand{\beq}{\begin{equation}}
\newcommand{\eeq}{\end{equation}}
\newcommand{\bea}{\begin{eqnarray}}
\newcommand{\eea}{\end{eqnarray}}

\renewcommand{\and}{{\qquad {\rm and} \qquad}}

\newcommand{\Res}{\mathop{\,\rm Res\,}}

\newcommand{\ii}{{\mathrm{i}}}

\newcommand{\Pint}{{\int\kern -1.em -\kern-.25em}}

\begin{document}
%=============================Page de titre===============
%\date{??}
%\author{Eynard}
%\title{Correlation functions for hermitian random matrices}
%\topmargin .5cm \textheight 21.5cm \textwidth 15.8cm 
%\oddsidemargin 0.54cm
%\evensidemargin 0.54cm 
\sloppy

%\maketitle

\pagestyle{empty}
\hfill preprint IPhT t17/139
\addtolength{\baselineskip}{0.20\baselineskip}
\begin{center}
\vspace{26pt}
{\large \bf {Large Strebel graphs and $(3,2)$ Liouville CFT}}\\
%\newline
\vspace{26pt}
S\'everin Charbonnier$^1$, Bertrand Eynard$^{1,2}$ and Fran\c cois David$^{3}$\\
\vspace{26pt}
\small{$^1$ Institut de physique th\'eorique, Universit\'e Paris Saclay, CEA, CNRS,\\ F-91191 Gif-sur-Yvette, France}\\
\small{$^2$ Centre de Recherches Math\'ematiques, Universit\'e de Montr\'eal, \\Montr\'eal, Canada}\\
\small{$^3$ Institut de physique th\'eorique, Universit\'e Paris Saclay, CNRS, CEA, \\F-91191 Gif-sur-Yvette, France}\\
\vspace{13pt}
\today
\end{center}

\vspace{20pt}
\begin{center}
{\bf Abstract}

2D quantum gravity is the idea that a set of discretized surfaces (called map, a graph on a surface), equipped with a graph measure, converges in the large size limit (large number of faces) to a conformal field theory (CFT), and in the simplest case to the simplest CFT known as pure gravity, also known as the gravity dressed (3,2) minimal model.
Here we consider the set of planar Strebel graphs (planar trivalent metric graphs) with fixed perimeter faces, with the measure product of Lebesgue measure of all edge lengths, submitted to the perimeter constraints.
We prove that expectation values of a large class of observables indeed converge towards the CFT amplitudes of the (3,2) minimal model.

\end{center}
%-----------------------------ABSTRACT--------------------------------------
%
%Abstract

%\begin{center}

%\end{center}

%\newpage
%\pagestyle{empty}

%\section*{}

%\newpage
\vspace{26pt}
\pagestyle{plain}
\setcounter{page}{1}

%\maketitle

\tableofcontents

\section{Introduction}

The idea of two-dimensional quantum gravity was born in the 1980's and developed in 1990's.
It consists in the study of two-dimensional (2d) surfaces equipped with a random Riemaniann metric.
By analogy with Euclidean path and functional integrals in quantum mechanics and quantum field theories and with general relativity, the randomness corresponds to quantization and the 2d metric to some 2d "gravitational field", whence the name "2d quantum gravity".

In 1981, motivated by string theory, Polyakov \cite{Polyakov81} argued that 2d-quantum gravity should be equivalent to a 2d-quantum field theory (CFT) on the surface, the Liouville conformal field theory (CFT). This theory has been extensively studied, see \cite{Nakayama2004} for an extensive but not too recent review.
Massless matter coupled to 2d gravity is described by a CFT characterized (sometimes uniquely) by its central charge $c$, and the central charge of the associated Liouville CFT should be $c_L=26-c$. (see for instance \cite{David1988-M}, \cite{DistlerKawai1989}). For pure gravity $c=0$.

Another idea is to discretize the problem: start from a set of discrete surfaces (also called random maps in the mathematical literature), for example triangulated surfaces with $N$ triangles, equipped with some local measure, for instance the uniform measure. These discretized models can often be mapped onto random matrix models, and have been studied by random matrix theory, by combinatorics and by statistical physics methods, including numerical methods. 
The continuum limit is defined by letting the average number of triangles tend to infinity, and the mesh to zero, while keeping the area fixed. 
It is conjectured that this limit should exist and be a 2d Liouville CFT. In order to identify the Liouville CFT, one has to embed the discrete surface on a surface with a metric, and measure expectation values and correlations of distances between points.
This limit is expected to be universal, in the sense that it should be the same for a large class of random maps (triangles, quadrangles, or other sets of graphs) rather independently of the measure.

In 1990 Di Francesco and Kutasov \cite{DiFrancescoKutasov1990} showed that for some special values of $c$, some quite special observables of the Liouville CFT (the partition functions and certain correlation functions) coincide with the amplitudes (coefficients of the $\tau$-function) of %degenerate Liouville theories 
an integrable system formulated by Douglas and Shenker \cite{DouglasShenker1990} as a reduction of the KdV integrable hierarchy, known as the $(p,q)$ minimal model. There is a central charge associated to this integrable system, given by $c=1-6(p-q)^2/pq$. 
For $c=0$, this degenerate Liouville theory is associated to the minimal model $(3,2)$. 
Therefore, in a setup supposed to be a discretized model of 2d pure quantum gravity, one should be able to relate the continuum limit to the minimal model $(3,2)$.

The purpose of this work is to establish this general equivalence for a specific discretization setup of pure 2d gravity, that we present now .
Consider an abstract triangulation in the plane (planar triangulation), and its dual, an abstract trivalent graph.
Many methods to embed such an abstract triangulation into the plane have been developed and studied, for example those based on circle packings \cite{Benjamini2010} and those on more general circle patterns \cite{DavidEynard2014}. One asset of these embeddings is that they show a conformal invariance even for a finite number of points, a property that one wants to have in the continuum limit, in order to make contact with CFT. 
The embedding that that we shall consider here is based on Strebel graphs \cite{HarerZagier1986}, \cite{Penner1987}, \cite{Kontsevich1992}. 
Strebel graphs are metric trivalent ribbon graphs drawn on surfaces. ``Metric'' means that a length is associated to each edge, and their dual are triangulations. 
Strebel's theorem says that the graph's metric can be uniquely extended to a metric on the whole surface, with the curvature localized at face centers. 
The set of Strebel graphs of genus $g$ with $N$ faces is isomorphic to the moduli space of (decorated) Riemann surfaces of genus $g$ with $n$ marked points (the face centers) decorated by $N$ real numbers (the face perimeters). This is a non--compact space since perimeters can be as large as desired.
%The lengths of edges (decorations) are to some extend redundant for our problem, and some constraints can be enforced on them, without changing the continuum limit, hence the universality class (using physicist's langage) of the model.
In this work, and restrain the set of Strebel graphs to the  subset of graphs with uniform fixed perimeters for the faces (see below for a precise definition), and we choose the Kontsevich measure on Strebel graphs, which is local.

The features of the Kontsevich measure along with this restriction will allow to compute the partition function of the Strebel graphs and the expectation values of all the observables which have a topological interpretation in 2d gravity. 
Using the knowledge of Kontsevich--Witten planar intersection numbers, we shall derive explicit expressions for these correlation functions, and we shall be able to compute explicitly (by saddle point approximation) their continuum limit, and show that they tend to the (3,2) minimal model amplitudes.

Moreover, by a Laplace transform, we shall show how to associate a spectral curve to this discrete model, and write it explicitly, in terms of Bessel functions.
The spectral curve is an object that encodes all the observables of the model, and it will depend on a single parameter $\mu$. 
The continuum limit $N\to\infty$ of the model will be shown to be equivalent to a limit where $\mu$ approaches a critical value $\mu_c$. 
We shall show that as $\mu\to\mu_c$ the spectral curve tends to a universal and simple spectral curve, which is nothing but the spectral of the integrable system corresponding to the $(3,2)$ minimal model of Douglas-Shenker. 
In this way, we show that the expectation values of all the topological observables of the model tend in the continuum limit to the amplitudes of the $(3,2)$ minimal model. 

Following the equivalence stated by Di Francesco and Kutasov (\cite{DiFrancescoKutasov1990}), this paper shows that considering Strebel graphs with uniform perimeters is a relevant discretization of 2d quantum gravity, which allows to recover it in the continuum limit.\\

This paper is organized as follows. First, we set the notations and recall the definitions of Strebel graphs. We describe the measure and its relation to the Chern-Class measure on Moduli space of Riemann surfaces, following Kontsevich \cite{Kontsevich1992}. We then restrict the model on uniform perimeters, which is specific to this paper. Last, we define the observables and their generating functions.

In a second part, the explicit computation of generating functions --made possible by the restriction on uniform perimeters-- is carried out. 

The third part is dedicated to the spectral curve and its critical form. It then contains the main result of this paper.

Last, in a fourth section, as an application, we derive the large size limit of several observables by different means (analysis of the partition function, saddle point method, and use of the spectral curve).

\section{Strebel graphs with uniform perimeters}
\subsection{General definitions -- Strebel graphs, moduli space of Riemann surfaces and Chern classes}
\subsubsection{Strebel and Kontsevich graphs}
\subsubsection*{Strebel graphs}

A  Strebel graph of genus $g$ with $n$ faces, is a trivalent cellular ribbon graph, that can be embedded on a surface of genus $g$, whose faces are topological discs, and whose edges $e$ carry a real positive number called the edge length $\ell_e\geq 0$. 
Strebel's theorem \cite{Strebel1984} provides a canonical embedding of the Strebel graph on a Riemann surface, equipped with a canonical metric, in such a way that each edge $e$ is a geodesic of length $\ell_e$ (see appendix \ref{appStrebel}).

We shall call:
\begin{itemize}
\item $\mathcal F$= set of faces
\item $\mathcal V$= set of vertices
\item $\mathcal E$= set of edges, and $\mathcal E_f$ the set of edges adjacent to a face $f$.
\end{itemize}

If a graph is planar, if we denote $N=|\mathcal F|-3$ ($|\mathcal F|$ denotes the cardinal), we have
\begin{equation}
|\mathcal E| = 3N+3
\qquad , \qquad
|\mathcal V| = 2N+2.
\end{equation}

\begin{figure}
\centering
\includegraphics[width=12.5cm]{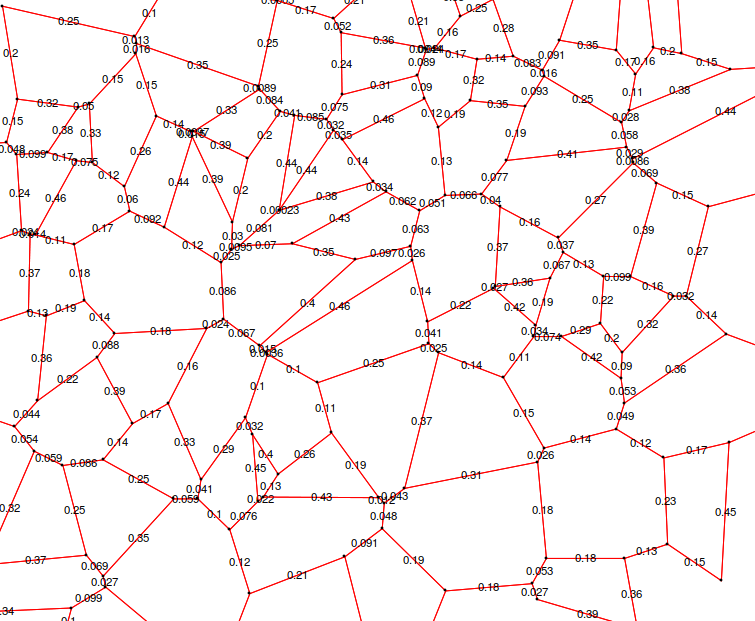}
\caption{Sample of Strebel Graph with $L_f=1$: the figure shows a sample from a Strebel graph. All the vertices are trivalent, and a positive real number is associated to each edge. Summing over the lengths of a face yields $L_f=1$.}
\label{strebel_graph}
\end{figure}

The face perimeters 
$$
L_f = \sum_{e\in \mathcal E_f} \ell_e
$$
play a special role. In figure \ref{strebel_graph}, a portion of a planar Strebel graph with all face perimeters equal to 1 ($L_f=1$) is represented.

%$$
%L_{\rm face}=\sum_{{\rm edge}\,\in\,{\rm edges\, adjacent\,to\,face}} \ell_{\rm edge}$$
%play a special role.

Kontsevich studied the set of Strebel graphs, equipped with the measure product of edge measures
\begin{equation}\label{defmeasureStrebel}
{\rm measure} = \prod_{e\in\mathcal E} d\ell_e \,\,\prod_{f\in \mathcal F} 
\delta(L_f - \sum_{e\in \mathcal E_f} \ell_e).
\end{equation}
In the planar case, we may chose an edge basis $\mathcal E_0\subset \mathcal E$ of cardinal $|\mathcal E_0|=2N$ (thus solving the perimeter constraints), and we also have (see \cite{Kontsevich1992})
\begin{equation}\label{defmeasureStrebelE0}
{\rm measure} = \frac12 \prod_{e\in\mathcal E_0} d\ell_e .
\end{equation}
This measure is not normalized, one of our goals will be to compute the total volume.

%
%Kontsevich also studied the Laplace transformed version
%\begin{equation}
%{\rm measure}' = \prod_{e\in\mathcal E} d\ell_e \,\,\prod_{f\in \mathcal F} 
%e^{-\lambda_f L_f},
%\end{equation}
%where $\lambda_f$ are "decoration parameters" associated to each faces.

\subsubsection*{Kontsevich graphs}

In fact Kontsevich was also interested in computing Laplace transforms of various observables, by Laplace transformation over the perimeters.
The Laplace transform with respect to perimeters, reformulates the problem in terms of the dual graphs. Since vertices of a Strebel graph are generically trivalent, its dual --that we shall call a "Kontsevich graph"-- is a triangulation of the surface, whose vertices $z_i$ are the center of faces of the Strebel graph.
Instead of carrying edge lengths, Kontsevich graphs carry a variable $\lambda_i$ at each vertex, and the positions $z_i$ of the vertices on the Riemann surface. 

Kontsevich studied the set of Strebel graphs with $|\mathcal F|=N+3$ vertices, equipped with the Chern-class measure
\begin{equation}\label{defmeasureK}
{\rm measure}_{\rm K} = \frac{1}{N!}\,\left(\sum_{i=1}^{N+3} L_i^2  \,c_1(T^*_{z_i}) \right)^{N+3g}
\end{equation}
where $c_1(T^*_{z_i})$ is a 2-form, the Chern class of the cotangent bundle $T^*_{z_i}$ at the $i^{\rm th}$ marked point $z_i$, and $g$ is the genus of the surface (we shall specialize to the planar case $g=0$).

He showed that this measure is in fact proportional to the measure \eqref{defmeasureStrebel} and \eqref{defmeasureK} by a power of $2$, that we shall recall below.

\subsubsection{Moduli spaces of surfaces}

Let $\mathcal M_{g,N}$ be the moduli space of Riemann surfaces of genus $g$, and with $N$ marked points:
\begin{equation}
\mathcal M_{g,N} = \{ (\Sigma, p_1,\dots,p_N)\}/{\rm Aut}
\end{equation}
where $\Sigma$ is a Riemann surface of genus $g$ and $p_1,\dots,p_N$ are $N$ distinct and labelled marked points on $\Sigma$. Two Riemann surfaces are isomorphic iff there is an analytic bijection (whose inverse is also analytic) that maps one to the other, respecting the marked points.
$\mathcal M_{g,N}$ is an orbifold (locally a manifold quotiented by a group -- the group of automorphisms), of real dimension
\begin{equation}
\dim \mathcal M_{g,N} = 2(3g-3+N).
\end{equation}
This means that it can be parametrized (locally) by $2(3g-3+N)$ real parameters, or also by $3g-3+N$ complex parameters.

From now on, we shall focus on the planar case $g=0$. We shall also consider that the number of marked points be $N+3$.
We have
\begin{equation}
\dim \mathcal M_{0,N+3} = 2N.
\end{equation}
Indeed, there is a unique (up to automorphisms) Riemann surface of genus 0, this is the Riemann sphere, i.e. the complex plane compactified by adding a point at $\infty$, and this is also the complex projective line, we write it
\begin{equation}
\bar{\mathbb C} = \mathbb C \cup \{\infty\} = \mathbb CP^1.
\end{equation}
Automorphisms of the Riemann sphere, are M\"obius transformations $z\to (az+b)/(cz+d)$ with $ad-bc=1$, i.e.
\begin{equation}
{\rm Aut}(\bar{\mathbb C}) = Sl_2(\mathbb C).
\end{equation}
This means that, by chosing $a,b,c,d$, one can map any 3 of the marked points, let's say $p_1,p_2,p_3$ to 3 given points, let's say $0,1,\infty$.
In other words an element $(\bar{\mathbb C}, p_1,\dots,p_{N+3})\in \mathcal M_{0,N+3}$ is equivalent to $(\bar{\mathbb C}, 0,1,\infty,p_4,\dots,p_{N+3})$, and thus to the data of $N$ distinct complex numbers $p_4,\dots,p_{N+3}$.
This shows that $\dim_{\mathbb C} \mathcal M_{0,N+3}=N$, and thus $\dim_{\mathbb R} \mathcal M_{0,N+3}=2N$.

\subsubsection*{Decoration with perimeters}

We shall now consider the space
\begin{equation}
\tilde{\mathcal M}_{0,N+3} = \mathcal M_{0,N+3}\times \mathbb R_+^{N+3},
\end{equation}
i.e. we associate a positive real number $L_i$ to each marked point $p_i$.
It is also a trivial real bundle over $\mathcal M_{0,N+3}$, with fiber $\mathbb R_+^{N+3}$.
It has dimension:
\begin{equation}
\dim_{\mathbb R} \tilde{\mathcal M}_{0,N+3}=2N+N+3=3N+3.
\end{equation}
Strebel, Penner, Harrer, Zaguier, Kontsevich found that
\begin{equation}\label{eqMgnStrebel}
\tilde{\mathcal M}_{0,N+3} \sim \oplus_{G\in \mathcal G_{0,N+3}} \mathbb R_+^{\mathcal E(G)}
\end{equation}
where $\mathcal G_{0,N+3}$ is the set of planar Strebel graphs with $N+3$ faces, and $\mathcal E(G)$ its set of edges.
The isomorphism is an orbifold--isomorphism, i.e. respecting the quotients by automorphism groups on both sides.
The $z_i$ are the centers of faces of Strebel graphs, i.e. vertices of Kontsevich graph, and the $L_i$s are the perimeters.
In other words, a point of $\tilde{\mathcal M}_{0,N+3}$, is uniquely represented by a Strebel graph (or its dual the Kontsevich graph which is a triangulation), and the edge lengths provide a set of real coordinates.

\subsubsection{Chern classes on moduli space of curves of genus 0.%$\mathcal M_{0,N+3}$
}

Let us consider the bundle $\mathcal L_i$ over $\mathcal M_{0,N+3}$, whose fiber over $(\bar{\mathbb C},z_1,\dots,z_{N+3})\in \mathcal M_{0,N+3}$ is the cotangent plane $T^*_{z_i}{\bar{\mathbb C}}$ of the Riemann sphere at the $i^{\rm th}$ marked point $z_i$:
\begin{align}
\mathcal L_i & \to  \mathcal M_{0,N+3} \cr
(\bar{\mathbb C},z_1,\dots,z_{N+3},T^*_{z_i}{\bar{\mathbb C}})  & \mapsto  (\bar{\mathbb C},z_1,\dots,z_{N+3}).
\end{align}
The fiber is homeomorphic to the complex plane, it is thus a complex line bundle.
Let us denote 
\begin{equation}
\psi_i=c_1(\mathcal L_i)
\end{equation} 
its 1st Chern class.
We also consider the bundle $\tilde{\mathcal L}_i$ over $\tilde{\mathcal M}_{0,N+3}$, whose fiber over $(\bar{\mathbb C},z_1,\dots,z_{N+3},L_1,\dots,L_{N+3})\in \tilde{\mathcal M}_{0,N+3}$ is again the cotangent plane $T^*_{z_i}{\bar{\mathbb C}}$ of the Riemann sphere at the $i^{\rm th}$ marked point $z_i$, and denote $\tilde \psi_i=c_1(\tilde{\mathcal L}_i)$ its 1st Chern class. 
Since $\tilde{\mathcal M}_{0,N+3}={\mathcal M}_{0,N+3}\times \mathbb R_+^{N+3}$ is a product bundle, the Chern classes add, and since $\mathbb R_+^{N+3}$ is a flat bundle its Chern class vanishes, so that we may identify
\begin{equation}
\psi_i = \tilde \psi_i.
\end{equation}
Kontsevich found that, in the edge lengths coordinates, the Chern class takes the form
\begin{equation}
\psi_i = \sum_{e<e' \,,\, {\rm adjacent\,to}\, z_i} d\left(\frac{\ell_e}{L_i}\right)\wedge d\left(\frac{\ell_{e'}}{L_i}\right)
\qquad , \quad
L_i = \sum_{e \,,\, {\rm adjacent\,to}\, z_i} \ell_e.
\end{equation}
This may seem to depend on a choice of labelling of edges around $z_i$ (i.e. choosing a first edge and then order edge labels counterclockwise), but one can easily check that it doesn't depend on which edge is chosen to be the first.

$\psi_i$ is a 2-form on $\mathcal M_{0,N+3}$, and therefore $\left( \sum_i L_i^2\psi_i \right)^N$ is a top dimensional volume form on $\mathcal M_{0,N+3}$, and multiplied by $\prod_i dL_i$ it is a top dimensional volume form on $\tilde{\mathcal M}_{0,N+3}$.
It is thus proportional to $\prod_e d\ell_e$, and Kontsevich found that
\begin{equation}\label{eqrelpsiell}
\left( \sum_i L_i^2\psi_i \right)^N\prod_i dL_i =
N!\,\,2^{2N+1} \prod_e d\ell_e.
\end{equation}

\subsection{Restriction of the model, definition of the observables}

2D quantum gravity requires to carry out averages over the set of all possible --conformally non-equivalent-- metrics on all possible compact complex surfaces. Actually, it is possible to restrain the sum over connected compact Riemann surfaces, and (using the conformal gauge fixing of \cite{Polyakov81}) the sum over the metrics is reduced to the sum over a local conformal factor (the Liouville field) and some moduli. To leading order in the topological expansion (the planar limit) one can consider only the fields living on a genus 0 surface (i.e. the Riemann sphere). We focus on this leading order in this paper. 
Yet, the measure over the Liouville field is not easy to construct (see however \cite{DavidetalLiouville2016} for a recent rigourous construction of this measure).  
Therefore, as in the standard discretization schemes, we shall approach the set of Liouville fields (which is an infinite dimensional space) by a sequence of finite dimensional spaces. These are precisely $\tilde{\mathcal{M}}_{0,N+3}$ and $\mathcal{G}_{0,N+3}$. Every point in these spaces, through Strebel's theorem is equivalent to a metric over the Riemann sphere with $N+3$ punctures. The hope is that the limit $N\to \infty$ gives the continuous 2D pure (i.e. without matter fields) quantum gravity. Actually, each Strebel graph (see appendix \ref{appStrebel}) represents a flat metric over the Riemann sphere with $N+3$ punctures, all the curvature being located at the vertices. It is thus a special class of metrics.

Moreover, we will restrain this to a certain subset of Strebel graphs, namely the graphs with uniform perimeters: $L_1=\dots=L_{N+3}=L$, so that the space of metrics is even more specific. 
%Yet, these metrics are expected to belong to the same universality class as the one of the general Strebel graphs. 
However, we expect universality, i.e.
that the large $N$ limit (the continuum limit) of the observables will be 
independent of the type of Strebel graphs considered. In the end, the sum over these particular metrics shall already yield the 2D pure quantum gravity. 
In other word, the additional variables (the perimeters of the face) should be irrelevant redundant variables which do not change the continuum limit.
\\

The aim of this paper is thus to compute the large $N$ limit of the following observables defined on the set of Strebel Graphs with fixed number of faces. These observables have to be understood as pure gravity correlation functions. They are averages of a variable over all possible metrics. In this section, we first define the observables. As the computation of the observables is an enumeration problem, we encode each observable in a generating function.
Then, using known results on moduli spaces, we give an explicit computation of the generating functions.

\subsubsection{Volume}

We are interested in the measure on the moduli space $\tilde{\mathcal M}_{0,N+3}$:
\begin{equation}
d\mu = \prod_e d\ell_e = \frac{1}{N! 2^{2N+1}}\,\left( \sum_i L_i^2\psi_i \right)^N  \prod_{i} dL_i.
\end{equation}
Its volume is clearly infinite, because the volume of the fiber $\mathbb R_+^{N+3}$ is infinite.
We may however compute the volume of a strata with fixed perimeters $\mathbf L=(L_1,\dots,L_{N+3})$
\begin{eqnarray}
\mathcal Z_{N+3}({\mathbf L}) 
&=&  \int_{\tilde{\mathcal M}_{0,N+3}(L_1,\dots,L_{N+3})} d\mu \cr
&=&  \frac{1}{N! 2^{2N+1}}\,\int_{{\mathcal M}_{0,N+3}} \left( \sum_i L_i^2\psi_i \right)^N  .
%&=&  \sum_{G\in \mathcal G_{0,N+3}} \frac{1}{|{\rm Aut}(G)|}\prod_{e\in \mathcal E} d\ell_e \,\,\prod_{i=1}^{N+3} \delta(L_i-\sum_{e\in \mathcal V_{z_i}} \ell_e).
\end{eqnarray}
By the Kontsevich's theorem, we can rewrite this volume over $\tilde{\mathcal M}_{0,N+3}$ as a volume over Strebel graphs with $N+3$ faces:
\begin{equation}
\mathcal Z_{N+3}({\mathbf L})= \sum_{G\in \mathcal G_{0,N+3}} \frac{1}{|{\rm Aut}(G)|}\prod_{e\in \mathcal E} d\ell_e \,\,\prod_{i=1}^{N+3} \delta(L_i-\sum_{e\in \mathcal V_{z_i}} \ell_e).
\end{equation}
It then corresponds to the volume of the set of Strebel graphs with $N+3$ faces, whose perimeters are $(L_1,\dots,L_{N+3})$. 
Let us simplify the formulae. First, on the Strebel graphs side, whenever the surface has marked points, there is no non-trivial automorphisms, $|{\rm Aut}(G)|=1$, so the volume is:
\begin{equation}
\mathcal Z_{N+3}({\mathbf L})= \sum_{G\in \mathcal G_{0,N+3}}\prod_{e\in \mathcal E} d\ell_e \,\,\prod_{i=1}^{N+3} \delta(L_i-\sum_{e\in \mathcal V_{z_i}} \ell_e).
\end{equation}
Second, on the moduli space side, the standard convention is that if a form is integrated on a cycle whose dimension is not equal to the form's dimension, then the integral is zero.
For example we may write here:
\begin{equation}
\frac{1}{2^{2N} N!}\int_{\mathcal M_{0,N+3}} (\sum_i L_i^2 \psi_i)^N
= \int_{\mathcal M_{0,N+3}} e^{\frac14\left(\sum_i L_i^2 \psi_i\right)}
= \frac{1}{2^N}\int_{\mathcal M_{0,N+3}} e^{\frac12\left(\sum_i L_i^2 \psi_i\right)}.
\end{equation}
We shall use the standard Witten's notation for powers of the Chern classes
\begin{equation}
\psi_i^d = \tau_d,
\end{equation}
and for the {intersection numbers} 
\begin{equation}
\left\langle \prod_{i=1}^k \tau_{d_i} \right\rangle_g = \int_{\mathcal M_{g,k}} \prod_i \tau_{d_i} = \int_{\mathcal M_{g,k}} \prod_i \psi_i^{d_i}
\end{equation}
which -- by convention --  are zero if $\sum_i d_i \neq 3g-3+k$. 
The volume of our moduli space is then
\begin{align}
2 \mathcal Z_{N+3} ({\mathbf L}) 
&=  \int_{\mathcal M_{0,N+3}} e^{\frac14 \sum_i L_i^2 \psi_i} \cr
&=  \int_{\mathcal M_{0,N+3}} \prod_{i=1}^{N+3} e^{\frac14 L_i^2 \psi_i} \cr
&= \int_{\mathcal M_{0,N+3}}  \prod_{i=1}^{N+3} \left(\sum_{d_i} \frac{L_i^{2d_i}}{2^{2d_i} d_i!}\,\tau_{d_i}\right) \cr
&=  \sum_{d_1+\dots +d_{N+3}=N}\,\prod_{i=1}^{N+3}  \frac{L_i^{2d_i}}{2^{2d_i} d_i!}\,\, \left\langle \tau_{d_1}\tau_{d_2} \dots \tau_{d_{N+3}} \right\rangle_0 \cr
\end{align}

Let us now consider Strebel graphs of genus 0 with the same fixed perimeter $L$ for all faces ($L_i =L$ for all $i$). We are then interested in the following volumes:
\begin{eqnarray}
    \mathcal Z_{N+3}(L) & \overset{\mathrm{def}}{=} &    \mathcal Z_{N+3}(\overbrace{L,\dots,L}^{N+3}) \cr
&=&    \int_{\mathcal{M}_{0,N+3}(L,\dots,L)}d\mu ,
\end{eqnarray}
which we can restate -- from what precedes -- as:
\begin{equation}
\mathcal Z_{N+3}(L)=4 \left\langle  \left(\frac12 \sum_d \frac{L^{2d}}{2^{d} d!}\,\tau_d\right)^{N+3} \right\rangle_0.
\end{equation}
 The \textbf{generating function} associated to the volume is defined as:
\begin{align}
\hat{\mathcal Z}(\mu,L) 
& = \sum_{N} \frac{\mu^{N+3}}{(N+3)!}\,\mathcal Z_{N+3}(L)  \cr
& = 4 \left< \exp{\left(\frac{\mu}{2} \sum_d \frac{L^{2d}}{2^d\, d!}\,\tau_d \right)} \right>_0 .
\end{align}
%The volume of possible metrics is not a proper observable in the sense that is not a quantity that one can observe (in the same manner as a partition function), but one has to compute it in order to carry out averages over metrics.
%The proper observables are defined in the next section.

\subsubsection{Correlation functions}
In the present model, all the faces of a graph have the same perimeter $L$. 
%As it has been known since Einstein, gravity and metric on a space are closely related. 
The perimeters of a genus 0 Strebel graph are the lengths of closed geodesics of the punctured Riemann sphere, computed with the Strebel's metric (see appendix \ref{appStrebel}). 
As the measure of the perimeter of a face is directly linked to the way the metric behaves in this face, and as the metric contains all the "gravitational" information, then the measure of the perimeter of a face shall be a "gravitational observable". We allow a finite set of faces to have a prescribed perimeter, that is to say, if we fix $n$,  we allow $n$ faces to have perimeters $L_1,\dots,L_n$. We then look at Strebel graphs with $N+3+n$ faces (here $n$ is fixed, and $N$ varies), $n$ of them having the prescribed perimeters $L_1,\dots,L_n$, and the $N+3$ others have perimeter $L$. Then, we define the following volumes for this kind of Strebel graphs:
\begin{eqnarray}
\mathcal{Z}_{N+3,n}(L;L_1,\dots,L_n)&\overset{\mathrm{def}}{=}&
    \mathcal Z_{N+3+n}(\overbrace{L,\dots,L}^{N+3},L_1,\dots,L_n) \cr
&=&    \int_{\mathcal{M}_{0,N+3+n}(L,\dots,L,L_1,\dots,L_n)}d\mu \cr
&=& 2^{2-n} \sum_{d_1,\dots,d_n} \left\langle  \left(\frac12 \sum_d \frac{L^{2d}}{2^{d} d!}\,\tau_d\right)^{N+3} \prod_{i=1}^n \frac{L_i^{2d_i}}{2^{d_i} d_i!} \tau_{d_i} \right\rangle_0.
\end{eqnarray}
The subsequent generating function is:
\begin{eqnarray}\label{generZ}
\hat{\mathcal Z}_n(\mu,L;L_1,\dots,L_n) 
&\overset{\mathrm{def}}{=}&   \sum_N \frac{\mu^{N+3}}{(N+3)!}  \mathcal Z_{N+3,n}(L;L_1,\dots,L_n)  \cr
&=& 2^{2-n} \sum_{d_1,\dots,d_n}  \left\langle  e^{\frac{\mu}2 \sum_d \frac{L^{2d}}{2^{d} d!}\,\tau_d
} \,\,\, \prod_{i=1}^n \frac{L_i^{2d_i}}{2^{d_i} d_i!} \tau_{d_i} \right\rangle_0.
\end{eqnarray}
Note that setting $n=0$, we recover the definition of the volumes, which is just a specification of these observables. 
We will need the auxiliary generating function $\mathcal{U}$, which does not take the lengths $L_i$ into account, defined as the Legendre transform of $\hat{\mathcal{Z}}_n$:
\begin{eqnarray}\label{generU}
&&{\mathcal U}_n(\mu,L;d_1,\dots,d_n)\cr
&=&  \frac12 \sum_N \frac{\mu^{N}\,L^{2D} }{2^{2(N-D)}\,(N+n)!}\,\,  \int_{\mathcal{M}_{0,N+3+n}}\left( \sum_d \frac{L^{2d}}{d!} \tau_d\right)^{N+3} \,\, \psi_{N+3+1}^{d_1} \dots \psi_{N+3+n}^{d_n} \cr
&=&\frac12 \sum_N \frac{\mu^{N}\,L^{2D}}{2^{2(N-D)}\,(N+n)!}\,\,  \sum_{\tilde d_1+\dots+\tilde d_{N+3} = N-D} \frac{L^{2(N-D)}}{\prod_{i=1}^{N+3} \tilde d_i!} \,\, \frac{(N+n)!}{\prod_{i=1}^{N+3} \tilde d_i! \prod_{i=1}^n d_i!}\cr
&=&  \frac12 \sum_N \mu^{N}\,L^{2N}\,\,\prod_{i=1}^n \frac{1}{d_i!} \sum_{\tilde d_1+\dots+\tilde d_{N+3} = N-D} \prod_{i=1}^{N+3}\frac{1}{ 2^{2\tilde d_i}\tilde d_i!^2}\,\, .
\end{eqnarray}
It is possible to compute these generating function explicitly, which will be done in the next section. One efficient method of computation is to use the Eynard-Orantin Topological Recursion. We will use it to get the large $N$ behaviour of the observables. The Topological Recursion requires to encode the generating functions in differential forms. These differential forms are defined on a "spectral curve". 
In order to get complex parameters likely to live on a spectral curve (or, more generally, allowing the use of complex analysis methods), we carry out the Laplace transform of the generating functions with respect to the $L_i$:
\begin{eqnarray}
{\mathcal F}_n(\mu,L;z_1,\dots,z_n) 
&\overset{\mathrm{def}}{=}& \int_{0}^\infty dL_1 \dots \int_0^\infty dL_n e^{-\sum_i z_i L_i}  \hat{\mathcal Z}_n(\mu,L;L_1,\dots,L_n)  \cr
&=& 2^{2-n}  \sum_{d_1,\dots,d_n}  \prod_{i=1}^n \frac{(2d_i-1)!!}{z_i^{2d_i+1}} \left\langle  e^{\frac{\mu}2 \sum_d \frac{L^{2d}}{2^{d} d!}\,\tau_d
} \,\,\, \prod_{i=1}^n  \tau_{d_i} \right\rangle_0.
\end{eqnarray}
The differential forms are then obtained by differentiating with respect to $z_1,\dots,z_n$:
\begin{eqnarray}
{\mathcal W}_n(\mu,L;z_1,\dots,z_n) 
&\overset{\mathrm{def}}{=}& d_{z_1} \dots d_{z_n}  {\mathcal F}_n(\mu,L;z_1,\dots,z_n)  \cr
&=& (-1)^n\, 2^{2-n}  \sum_{d_1,\dots,d_n}  \prod_{i=1}^n \frac{(2d_i+1)!!\,dz_i}{z_i^{2d_i+2}} \left\langle  e^{\frac{\mu}2 \sum_d \frac{L^{2d}}{2^{d} d!}\,\tau_d
} \,\prod_{i=1}^n  \tau_{d_i} \right\rangle_0.\cr
\end{eqnarray}

\section{Explicit computations of generating functions}
It is possible to compute the correlation functions explicitly by taking advantage of the knowledge of the intersection numbers in genus 0. Indeed, the genus zero intersection numbers are (see for example \cite{Kontsevich1992}):
\begin{equation}\label{interg0}
<\tau_{d_1}\dots \tau_{d_{N+3}}>_0
= \frac{N!}{\prod_i d_i!} \,\,\delta_{N,\sum_i d_i}.
\end{equation}
The whole section relies on this result.

\subsection{Volumes of the strata}
\label{computeVolume}
It is easier to compute the 3rd derivative of the volume generating function. 
%We could have defined the generating function as the 3rd derivative of $\tilde{\mathcal{Z}}$, but it is less canonical (when it comes to the Topological ZRecursion) than the one we chose. 
Using \ref{interg0}, we get:
\begin{eqnarray}
\frac{\partial^3}{\partial \mu^3} \hat{\mathcal Z}(\mu,L)
& = & \sum_{N} \frac{\mu^{N}}{N!}\,\mathcal Z_{N+3}(L)  \cr
& = & \frac12 \sum_{N} \frac{\mu^{N}}{2^{2N}\,\,N!}\,\sum_{d_1+\dots+d_{N+3}=N} \frac{L^{2N}}{\prod_i d_i!} \left\langle \prod_i \tau_{d_i} \right\rangle_0 \cr
& = & \frac12 \sum_{N} \frac{\mu^{N}}{2^{2N}\,\,N!}\,\sum_{d_1+\dots+d_{N+3}=N} \frac{L^{2N}}{\prod_i d_i!}\,\,\frac{N!}{\prod_i d_i!} \cr
& = & \frac12 \sum_{N} \mu^N L^{2N}\sum_{d_1+\dots+d_{N+3}=N} \frac{1}{\prod_i 2^{2d_i} d_i!^2} \cr
\end{eqnarray}
Let us consider the first kind modified Bessel function $I_0(z)$:
\begin{equation}
I_0(z) = \sum_{d=0}^\infty \frac{z^{2d}}{2^{2d}\,d!^2}.
\end{equation}
We have
\begin{equation}
\sum_{d_1+\dots+d_{N+3}=N} \frac{1}{\prod_i 2^{2d_i} d_i!^2}
= [z^{2N}] I_0(z)^{N+3} = \operatorname{Res}_{z\to 0} \frac{dz}{z^{2N+1}} I_0(z)^{N+3}.
\end{equation}
where $[z^k]f(z)$ stands for the coefficient of $z^k$ in the expansion of $f$ around 0. \\
Therefore
\begin{eqnarray}
\frac{\partial^3}{\partial \mu^3} \hat{\mathcal Z}(\mu,L)
& = & \frac12 \sum_{N} \operatorname{Res}_{z\to 0} \frac{dz}{z^{2N+1}} I_0(z)^{N+3} (\mu L^2)^N \cr
& = & \frac1{4\pi\ii} \oint_{C} \frac{dz}{z}\,\frac{ I_0(z)^{3}}{1-\mu L^2 I_0(z)/z^2} 
\end{eqnarray}
where $C$ is the integration contour of fig.\ref{pole} below. Indeed, since $\hat{\mathcal Z}$ is a $\mu$ formal series, instead of surrounding only $0$, the integration contour of $z$ must surround all poles that tend to $0$ as $\mu\to 0$, and thus $C$ has to surround $\pm u(\mu L^2)$ defined as the $O(\mu)$ solution of
\begin{equation}\label{defu}
\mu L^2 =\frac{u^2}{I_0(u)}.
\end{equation}
The function $u^2/I_0(u)$ is plotted in fig.\ref{bessel} below.
 \begin{figure}[!ht]
    \centering
         \begin{minipage}[t]{9cm}
             \centering
             \includegraphics[width=7cm]{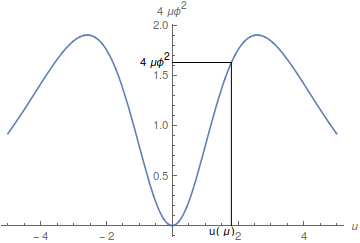}
             \caption{Plot of the function $\frac{u^2}{I_0 (u)}$}
             \label{bessel}
        \end{minipage}
 \begin{minipage}[t]{6cm}
 \centering
 \includegraphics[width=4cm]{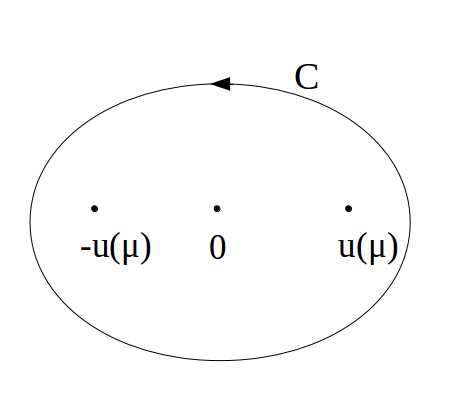}
 \caption{Contour of integration}
 \label{pole}
 \end{minipage}
 \end{figure}
 
The contour integral can be evaluated, it consists of residues of the 2 poles at $z=\pm u(\mu L^2)$:
\begin{eqnarray}
\frac{\partial^3}{\partial \mu^3} \hat{\mathcal Z}(\mu,L)
& = & \frac{1}{\mu L^2 u}\,\,\frac{ I_0(u)^{3}}{2 I_0(u)/u^3 - I_0'(u)/u^2}  \cr
%& = & \frac{u^2}{\mu L^2}\,\,\frac{ I_0(u)^{3}}{2 I_0(u) - u I_0'(u)}  \cr
& = & \frac{ I_0(u)^{4}}{2 I_0(u) - u I_0'(u)}  
\end{eqnarray}
The derivative of the Bessel function $I_0$ is the Bessel function $I_1$, thus 
\begin{equation}
\label{formulaVolume}
\frac{\partial^3}{\partial \mu^3} \hat{\mathcal Z}(\mu,L)
= \frac{ I_0(u)^{3}}{2 - u I_1(u)/I_0(u)}  
= \frac{u I_0(u)^2}{L^2}\,\,\frac{d u}{d\mu}.
\end{equation}
Using $I_1' = I_0 - I_1/u$, we can integrate:
\begin{eqnarray}
\frac{\partial^2}{\partial \mu^2} \hat{\mathcal Z}(\mu,L)
& = & \frac{u^2( I_0(u)^2-I_1(u)^2)}{2 L^2}.
\end{eqnarray}
Further integration is not doable explicitly, but this formula fits for our purpose of getting the large $N$ volumes. 

\subsection{Correlation functions}

Let us fix $d_1,\dots,d_n$, and note $D\overset{\mathrm{def}}{=}\sum_i (d_i-1)$. We begin with the auxiliary generating function $\mathcal{U}$.
In the same manner as for the volumes, we introduce the Bessel function $I_0(z)$ and we have:
\begin{eqnarray}
{\mathcal U}_n(\mu,L;d_1,\dots,d_n)
&=&  \frac12 \sum_N \frac{\mu^{N}\,L^{2N}}{\prod_{i=1}^n d_i!}
\operatorname{Res}_{z\to 0} \frac{dz}{z^{1+2(N-D)}} I_0(z)^{N+3} \cr
&=&  \frac{1}{4\pi\ii} \,\frac{1}{ \prod_i d_i!} \oint_C \frac{z^{2D} dz}{z}\, \frac{I_0(z)^3}{1-\mu L^2 I_0(z)/z^2} \cr
\end{eqnarray}
The residue can be evaluated easily, at the two poles $z=\pm u(\mu L^2)$.
Beside, if $D<0$, there can be another pole at $z=0$.
We have
\begin{eqnarray}
{\mathcal U}_n(\mu,L;d_1,\dots,d_n)
&=&  \frac{1}{\prod_i d_i!} \,  \frac{u^{2D}\,I_0(u)^4}{2 I_0(u)-u I_1(u)} \cr
&& +  \frac{1}{2} \,\frac{1}{\prod_i d_i!} \operatorname{Res}_{z\to 0} z^{2D+1} dz\, \frac{I_0(z)^3}{z^2-\mu L^2 I_0(z)} \cr
&=&  \frac{1}{\prod_i  d_i!} \,  \frac{u^{2D}\,I_0(u)^4}{2 I_0(u)-u I_1(u)} \cr
&& -  \frac{1}{2} \,\frac{1}{\prod_i d_i!} \sum_{j=1}^{-D} \frac{1}{(\mu L^2)^{j}} \operatorname{Res}_{z\to 0} z^{2(D+j)-1} dz\, I_0(z)^{3-j} .\cr
\end{eqnarray}

If $D<0$, the first term, proportional to $u^{2D}$ is a Laurent formal series of $\mu L^2$, starting with a negative power, whereas the last term contributing only if $D<0$,  is a polynomial of $1/\mu L^2$.
Since the whole result should be a power series of $\mu L^2$ with only positive powers, we understand that the last term just cancels the negative part of the first. We thus may write:
\begin{equation}
{\mathcal U}_n(\mu,L;d_1,\dots,d_n) 
=  \frac{1}{\prod_i  d_i!} \,\left(  \frac{u^{2D}\,I_0(u)^4}{2 I_0(u)-u I_1(u)} \right)_+
\end{equation}
meaning that we keep only positive powers of $\mu L^2$ in the Laurent expansion.
We observe that upon multiplying by $\prod_i d_i!$, the right hand side depends only on $D$ and $u$, we write it
\begin{equation}
{\mathcal U}_n(\mu,L;d_1,\dots,d_n) 
=  \frac{1}{\prod_i  d_i!} \,f_D(u)
\quad , \quad f_D(u)=
\left(  \frac{u^{2D}\,I_0(u)^4}{2 I_0(u)-u I_1(u)} \right)_+ .
\end{equation}

The relationship to our previously defined generating function is
\begin{eqnarray}
\hat{\mathcal Z}_n(\mu,L;L_1,\dots,L_n)
&=& L^{2n} \sum_{d_1,\dots,d_n} \prod_{i=1}^n \frac{L_i^{2 d_i} L^{-2d_i}}{2^{2d_i} d_i!} \,\,\partial_\mu^{n-3} \left( \mu^n {\mathcal U}_n(\mu,L;d_1,\dots,d_n)  \right)
\end{eqnarray}
So
\begin{eqnarray}
&&\hat{\mathcal Z}_n(\mu,L;L_1,\dots,L_n)\cr
&=& L^{2n} \sum_{d_1,\dots,d_n} \prod_{i=1}^n \frac{L_i^{2 d_i} L^{-2d_i}}{2^{2d_i} d_i!^2} \,\,\partial_\mu^{n-3} \left( \mu^n f_D(u(\mu L^2))  \right) \cr
&=& \partial_\mu^{n-3} \left( \mu^n L^{2n}  \sum_D   f_D(u(\mu L^2))   \sum_{d_1+\dots+d_n=D+n} \prod_{i=1}^n \frac{L_i^{2 d_i} L^{-2d_i} }{2^{2d_i} d_i!^2} \,\right) \cr
&=& \partial_\mu^{n-3} \left( \mu^n L^{2n}  \sum_D   f_D(u(\mu L^2))  
\operatorname{Res}_{z\to 0} \frac{dz}{z^{1+2(D+n)}}\,  \prod_{i=1}^n  I_0(z L_i/L) \,\right) \cr
&=& \partial_\mu^{n-3} \left( \mu^n L^{2n} 
\operatorname{Res}_{z\to 0} \frac{dz}{z^{1+2n}}\,  \prod_{i=1}^n  I_0(z L_i/L) \, \sum_{D=-n}^\infty  z^{-2D}\, f_D(u(\mu L^2))   \right) \cr
\end{eqnarray}

Carrying the sum over $D$ is possible if we impose $|z|>|u|$, so enforcing this condition, one gets:
\begin{equation}
\frac{1}{z^{1+2n}}\sum_{D=-n}^{+\infty}\frac{f_D(u(\mu L^2))}{z^{2D}}=\left( \frac{1}{u^{2n}}\frac{z}{z^2-u^2}\frac{I_0(u)^4}{2I_0(u)-uI_1(u)}    \right)_+
\end{equation}
Then we can rewrite $\hat{\mathcal{Z}}$ in the following way:
\begin{eqnarray}
\hat{\mathcal Z}_n(\mu,L;L_1,\dots,L_n)
&=& \frac{1}{2\pi\ii} \, \partial_\mu^{n-3} \left( \mu^n L^{2n}\,\left[ \frac{1}{u^{2n}}\frac{I_0(u)^4}{2I_0(u)-uI_1(u)}\times \right.\right.\cr
& &\left.\left. \oint_{C} \frac{z dz}{z^2-u^2}\,  \prod_{i=1}^n  I_0(z L_i/L) \, \, \right]_+ \,\,   \right) \cr
&=&  \partial_\mu^{n-3} \left( \mu^n L^{2n}\,\left[ \frac{1}{u^{2n}} \frac{I_0(u)^4}{2I_0(u)-uI_1(u)}
  \prod_{i=1}^n  I_0(u L_i/L) \, \, \right]_+ \,\,   \right) \cr
\end{eqnarray}
The contour integral is now $C$ (see figure \ref{pole}), because, though the residue is around $0$, we imposed $|z|>|u|$, in order to sum over $D$. 
Its Laplace transform is
\begin{eqnarray}
\mathcal F_n(\mu,L;z_1,\dots,z_n)
=  \partial_\mu^{n-3} \left( \mu^n L^{2n}\,\left[\frac{1}{u^{2n}} \frac{I_0(u)^4}{2I_0(u)-uI_1(u)}
  \prod_{i=1}^n  (z_i^2-u^2/L^2)^{-1/2}  \right]_+ \right). \cr
\end{eqnarray}
Again, note that the third derivative simplifies the result:
\begin{equation}
\partial_\mu^3 \mathcal{F}_n(\mu,L;z_1,\dots,z_n)
=  \partial_\mu^{n} \left( \frac{\mu^n L^{2n}}{u^{2n}}\, \frac{I_0(u)^4}{2I_0(u)-uI_1(u)}
  \prod_{i=1}^n  (z_i^2-u^2/L^2)^{-1/2} \,\,   \right).
\end{equation}

\section{Spectral curve}
All the combinatorics of the Strebel graphs is encoded in one complex curve: the Spectral Curve. It is the main object needed to run the Topological Recursion. Here, we use the fact that Topological Recursion solves the combinatorics of Strebel graphs and allows to compute all correlation functions. The first step is to determine the Spectral Curve. Actually, we find a family of Spectral Curves, indexed by the parameter $\mu$ introduced in the previous part. We first give the generic form of the Spectral Curve, and then a singular curve obtained when the parameter $\mu$ approaches a singular value $\mu_c$.
\subsection{Generic Spectral Curve}
One can re--express the generating function $\hat {\mathcal Z}(\mu,L)$ in the following way:
\begin{equation}
\, \hat{\mathcal Z}(\mu,L)
= \,4\,\left\langle \,e^{\frac{\mu}{2}\sum_d \frac{L^{2d}}{2^{d}\,d!} \,\tau_d} \right\rangle_0
\end{equation}
and similarly
\begin{equation}
2^{n-2-\sum_i d_i} L^{-2D} \frac{\partial^{n-3}}{\partial \mu^{n-3}} \mu^n\,{\mathcal U}_n(\mu,L,d_1,\dots,d_n)
=  \,\left\langle \tau_{d_1}\dots\tau_{d_n} \,e^{\frac{\mu}{2}\sum_d \frac{L^{2d}}{2^{d}\,d!}\,\tau_d} \right\rangle_0
\end{equation}
Kontsevich proved (this was Witten's conjecture) that
\begin{equation}
\left\langle \,e^{\frac12 \sum_d (2d-1)!!\,t_{2d+1} \,\tau_d} \right\rangle_{\rm all\,genus}
= \mathcal T_{\rm KdV}(\frac12 (2d-1)!! t_{2d+1})
\end{equation}
is the KdV-Tau function of the times $t_{2d+1}$s.
In other words, our generating function is equal to genus zero part of the KdV tau function evaluated at times
\begin{equation}
t_{2d+1} = \frac{\mu L^{2d}}{(2d)!}.
\end{equation}
Since the KdV tau function is independent of even times, we may chose
\begin{equation}
t_{k+1} = \frac{\mu L^{k}}{k !}.
\end{equation}

Beside, in \cite{Eynard2007}, \cite{Eynard2011}, it was shown that the following
\begin{equation}
\left\langle \,e^{\frac12 \sum_d (2d-1)!!\,t_{2d+1} \,\tau_d} \right\rangle
= e^{\sum_g F_g }
\end{equation}
where $F_g$s are the EO-invariants (defined in \cite{EynardOrantin2007}) of the spectral curve
\begin{equation*}
\mathcal S = 
    \left\{
    \begin{array}{ll}
         & x = z^2+\check{t}_1\\
         & y = z-\frac{1}{2}\sum_{k=0}^{+\infty}\check{t}_{2k+3}z^{2k+1}
    \end{array}
    \right.
\end{equation*}
with the coefficients $\check t_k$ related to the $t_k$s as follows:
\begin{equation*}
\check{t}_1 = \sum_{j=0}^{\infty} \frac{(2j-1)!!}{2^j j!} \check{t}^j_1  t_{2j+1}
\qquad , \qquad \check{t}_{2k+1} 
= \sum_{j=0}^{\infty} \frac{(2k+2j-1)!!}{(2k-1)!!\, 2^j j!} \check{t}^j_1  t_{2k+2j+1} .
\end{equation*}
In our case the equation determining $\check t_1$ is
\begin{equation*}
\check{t}_1 = \mu \sum_{j=0}^{\infty} \frac{1}{2^{2j} j! j!} \check{t}^j_1 L^{2j} 
= \mu I_0(L\sqrt{\check t_1}),
\end{equation*}
whose solution is
\begin{equation}
L \sqrt{\check t_1} = u(\mu L^2)
\end{equation}
with the function $u$ already introduced in \eqref{defu}.
And for higher times
\begin{eqnarray}
\check{t}_{2k+1} 
&=& \sum_{j=0}^{\infty} \frac{(2k+2j-1)!!}{(2k-1)!!\, 2^j j!} \check{t}^j_1  t_{2k+2j+1} \cr
&=& \frac{\mu L^{2k}}{(2k-1)!!} \sum_{j=0}^{\infty} \frac{1}{ 2^{2j+k} (k+j)! j!} \check{t}^j_1 L^{2j} \cr
&=& \frac{\mu L^{2k}}{(2k-1)!! u^k} \sum_{j=0}^{\infty} \frac{u^{2j+k}}{ 2^{2j+k} (k+j)! j!}  \cr
&=& \frac{\mu L^{2k}}{(2k-1)!! u^k} I_k(u) .
\end{eqnarray}
In other words, up to combinatorial prefactors, the spectral--curve times, are the Bessel functions of $u$.
\begin{equation}
\mathcal S = 
    \left\{
    \begin{array}{ll}
         & x = z^2+\frac{u^2}{L^2} \\
         & y = z-\frac{\mu}{2} \sum_{k=1}^{+\infty} \frac{L^{2k} I_k(u)} {(2k-1)!!\,u^k} \,\,z^{2k-1}
    \end{array}
    \right.
\end{equation}
In the spectral curve, $z$ is only a parameter, and reparametrizing
\begin{equation}
z= \frac{\sqrt{u}}{L}\,\zeta,
\end{equation}
we write the spectral curve as
\begin{equation}
\mathcal S = 
    \left\{
    \begin{array}{ll}
         & x = \frac{u}{L^2}\,\left(\zeta^2 + u\right) \\
         & y = \frac{\sqrt{u}}{L}\,\left( \zeta -\frac{u}{2 I_0(u)} \sum_{k=1}^{+\infty} \frac{I_k(u)} {(2k-1)!!} \,\,\zeta^{2k-1} \right)
    \end{array}
    \right.
\end{equation}

\subsubsection*{The one-form $ydx$}
The expression of the 1-form $ydx$ is:
\begin{equation}
ydx = \frac{u^{3/2}}{L^3}\,\left( 2\zeta^2 -\frac{u}{ I_0(u)} \sum_{k=1}^{+\infty} \frac{I_k(u)} {(2k-1)!!} \,\,\zeta^{2k} \right)\,d\zeta,
\end{equation}
which yields the derivative with respect to $\mu$:
\begin{eqnarray}
\left.\frac{\partial ydx}{\partial u}\right|_{{\rm fixed}\,x}
&=& \frac{-2u dz}{L^2} + \frac{\mu dz}{2} \sum_{k=1}^{+\infty} \frac{L^{2k} z^{2k}}{u^k\,(2k-1)!!}\,\left(\overbrace{I_{k+1}-I_{k-1}+2k\frac{I_k}{u}}^{= 0}\right)+\mu I_1 (u) dz \cr &=& \frac{-2u dz}{L^2} +\mu I_1 (u) dz. \cr
\end{eqnarray}

Its Laplace transform is;
\begin{eqnarray}
\int_{0}^{+\infty} ydx e^{-vx}
&=& \frac{1}{2}\frac{u^{3/2}}{L^3 \,I_0(u)}\,e^{-\frac{v u^2}{L^2} } \int_{\mathbb R} \left( 2 I_0(u) \zeta^2 - u \sum_{k=1}^{+\infty} \frac{I_k(u)} {(2k-1)!!} \,\,\zeta^{2k} \right) \, e^{-\frac{v u}{L^2} \zeta^2} \,d\zeta\cr
&=& \frac{1}{2}\frac{u \sqrt{\pi}}{L^2 \,I_0(u)\,\sqrt{v}}\,e^{-\frac{v u^2}{L^2} }\,  \left( \frac{L^2}{uv} I_0(u)  - u \sum_{k=1}^{+\infty} \frac{I_k(u) L^{2k}} {2^k u^k  v^k} \,\, \right)  \cr
&=& \frac{1}{2}\frac{\sqrt{\pi}}{I_0(u)}\,\frac{e^{-\frac{ v u^2}{L^2} }}{v^{3/2}}\,  \left( I_0(u)  - \frac{u^2 v}{L^2} \sum_{k=1}^{+\infty} \frac{I_k(u) L^{2k}} {2^k u^k v^{k} } \,\, \right)  \cr
\end{eqnarray}

%The generating function of Bessel functions is (in $\mathbb Q[s,1/s][[u]]$)
%\begin{equation}
%I_0 +  \sum_{k=1}^\infty (s^k+s^{-k}) I_k(u) = %\sum_{k=-\infty}^\infty s^k I_k(u)  = e^{u/2(s+1/s)}
%\end{equation}
%This gives
%\begin{eqnarray}
%\int ydx e^{-vx}
%&=& \frac{1}{2}\frac{\sqrt{\pi}}{I_0(u)}\,\frac{e^{-\frac{ v u^2}{L^2} }}{v^{3/2}}\,  \left( I_0(u)  - \frac{u^2 v}{L^2} ( e^{vu^2/L^2} e^{L^2/4v} - I_0 - \sum_k 2^k u^k v^k L^{-2k} I_k)\,\, \right)  \cr
%\end{eqnarray}

\subsection{Critical Spectral Curve}
\begin{figure}
\begin{minipage}[t]{14cm}
 \centering
 \includegraphics[width=13.5cm]{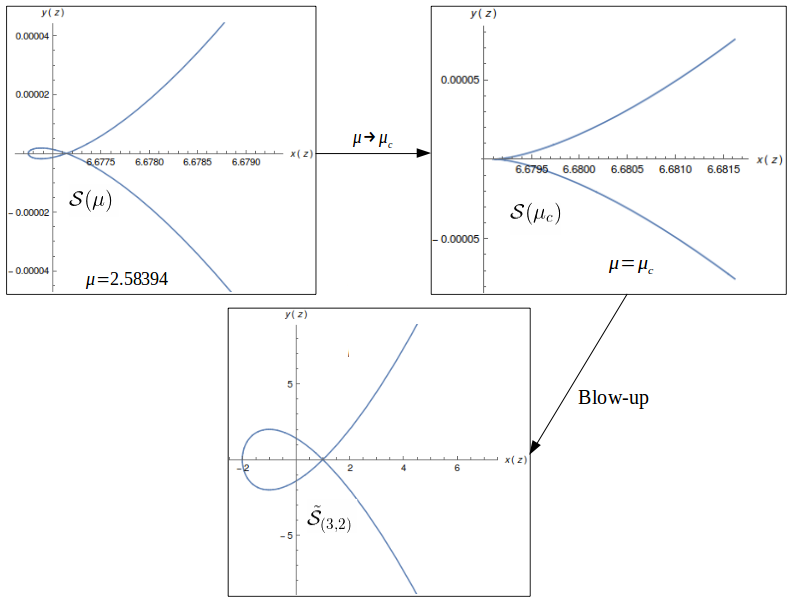}
 \caption{Projection in the plane $(x,y)\in\mathbb{R}^2$ of Spectral curves for different values of the index $\mu$. The second curve is the critical Spectral curve and shows the cusp. The third one is the curve near the critical point. At the leading order in $\mu-\mu_c$, it is the critical curve $\tilde{\mathcal{S}}_{(3,2)}$ of the minimal model $(3,2)$.}
 \label{parametricPlot}
 \end{minipage}
\end{figure}

Studying the large $N$ limit of a combinatorial data encoded in a generating function is closely related to the behaviour of the generating function near a singular point. The generating functions defined in the previous part all depend on the parameter $\mu$, on which depends the Spectral Curves $\mathcal{S}(\mu)$. When $\mu$ is close to $\mu_c$, a singular point of the generating functions, the Spectral curve $ \mathcal{S}(\mu)$ is close to a critical Spectral curve ${\mathcal{S}}(\mu_c)$ that is singular. This critical Spectral curve must provide the large $N$ behaviours of observables of the Strebel Graphs.\\
For a generic $\mu$, we have $2I_0(u(\mu))-u(\mu)I_1(u(\mu))\neq0$. Indeed, this quantity is null when $u(\mu)=u(\mu_c)=u_c$. We define the critical point $\mu_c$ by:
\begin{equation}
\mu_c = \frac{1}{L^2}\,\operatorname{max} \frac{u^2}{I_0(u)},
\end{equation}
this is the maximum of the curve in figure \ref{bessel}. 
The two preimages of $\mu_c$ are $u_c$ (say $u_c>0$) and $-u_c$. They satisfy:
\begin{equation}
\pm u_cI_1(\pm u_c)-2I_0(\pm u_c)=0.
\end{equation}
Their numerical values are: 
\begin{equation}
\pm u_c =\pm 2.5844...
\qquad ; \qquad
L^2 \mu_c = 1.902...
\end{equation}
Therefore, for $\mu\neq\mu_c$, that is $u\neq \pm u_c$, we have $y'(0) = 1-\frac{u I_1(u)}{2I_0(u)}\neq0$. The spectral curve $\mathcal{S}(\mu)$ is then  regular for $\mu<\mu_c$, and close to $\zeta=0$, $y$ behaves like $\sqrt{x-u^2/L}$. 
At $\mu=\mu_c$, $y$ behaves as a cusp $y\sim (x-u^2/L)^{3/2}$, and $\mathcal{S}(\mu_c)$ is no longer a regular spectral curve (see figure \ref{parametricPlot}).
Its Eynard-Orantin invariants diverge (see \cite{EynardOrantin2007}).
How they diverge is controlled by computing the resolution of the singularity, the blow up of the spectral curve in the vicinity of $\mu=\mu_c$.
Therefore, when $\mu\to \mu_c$, and thus $u\to u_c$, we rescale the variable $z$ as
\begin{equation}
z= \frac{\sqrt{u}}{L} \zeta  = -\sqrt{u_c-u} \,\frac{\sqrt{u}}{L}\,\xi.
\end{equation}
In that limit the spectral curve becomes
\begin{equation}
x = \frac{u_c^2}{L^2} + (u_c-u)\,\frac{u_c}{L^2} \,(\xi^2-2) + O((u_c-u)^2)
\end{equation}
\begin{equation}
y =  (u_c-u)^{3/2}\,\frac{u_c^2-4}{6L\sqrt{u_c}} \,(\xi^3-3\xi) + O((u_c-u)^2)
\end{equation}
The blow up of the spectral curve near its singularity is the rescaled curve
\begin{equation}
{\mathcal S}_{(3,2)} =
\left\{\begin{array}{l}
\tilde x(\xi) = \xi^2-2 \cr
\tilde y(\xi) = \xi^3-3\xi \cr
\end{array}\right.
\end{equation}
It is known \cite{Eynard2016} that this is the spectral curve of the $(3,2)$ minimal model, which according to \cite{DouglasShenker1990}, \cite{DiFrancescoKutasov1990} is equivalent to Liouville gravity with matter central charge $c=0$.

\section{Large N limits}

It remains to study the large $N$ behaviour of the observables.
Since $N+3+n$ is the number of  faces of the Strebel graph (number of vertices of the dual triangulation), the large $N$ limit should be the continuum limit of large maps, it should tend towards the Brownian map (according to \cite{LeGall2013}-\cite{Miermont2013}) and it is expected to converge towards Liouville theory.

As was mentionned in the previous part, large $N$ expansions are controlled by the singularities of the generating functions, that is to say we have to study the behavior as $\mu L^2 \to \mu_c L^2$, where $\mu_c $ is a point (closest to $0$) at which the generating functions are not analytic.

Volume and correlation functions large $N$ asymptotics are then related to the singular behavior of their respective generating functions when approaching the critical point $\mu_c$.

We first dwell on asymptotics of the volume, using the explicit computation we did in the second part. In order to compute the one point function at large $N$, we enforce the saddle point method in a second time. This allows us to identify a typical length scale for large maps. Last, we use Topological Recursion results and we use the critical Spectral curve to compute $n$-point functions.

\subsection{Asymptotics of the volume}
The third derivative of the generating function for the volume is given by formula \ref{formulaVolume} of section \ref{computeVolume}:
\begin{equation}
\partial_\mu^3 \hat{\mathcal{Z}}(\mu,L)=\frac{ I_0(u)^{4}}{2 I_0(u) - u I_1(u)}
\end{equation}
The critical point $\mu_c$, is the same as for the Spectral curve. Indeed, if $\mu=\mu_c$, one gets $2I_0(u(\mu_c))-u(\mu_c)I_1(u(\mu_c))=0$, so when $\mu\to\mu_c$, $\partial_\mu^3 \hat{\mathcal{Z}}$ diverges.

If $\mu$ is close to $\mu_c$, i.e. $u$ close to $u_c$,  we have:
 \begin{equation*}
\frac{\mu}{\mu_c}= 1- \frac{u_c^2-4}{2 u_c^2}(u_c-u)^2+O\left((u_c-u)^3\right)
\qquad , \qquad
\frac{u_c^2-4}{2 u_c^2} = 0.2005...
 \end{equation*}
i.e.
\begin{equation}
u_c-u \sim \sqrt{\frac{2u_c^2}{u_c^2-4}}\,\sqrt{1-\frac{\mu}{\mu_c} } \, (1+O(\sqrt{1-\mu/\mu_c}))
\qquad , \qquad
\sqrt{\frac{2u_c^2}{u_c^2-4}} = 2.23...
\end{equation}
 
So we get:
\begin{equation}
\partial_\mu^3\hat{\mathcal Z}(\mu,L) \underset{\mu\to\mu_c}{\sim}  \frac{C}{\sqrt{1-\frac{\mu}{\mu_c}}}\,+O(1)
\qquad , \qquad
C = \frac{1}{\sqrt{2}} \,\frac{I_0(u_c)^3}{\sqrt{u_c^2-4}} = 18.69...
\end{equation} 

$\partial_\mu^{3}\hat{\mathcal{Z}}$ behaves as $(1-\mu/\mu_c)^{-1/2}$, so  $\hat{\mathcal Z}(\mu,L)$ has a $(1-\mu/\mu_c)^{5/2}$ singularity.
 Writing that 
\begin{equation}
\frac{C}{\sqrt{1-\frac{\mu}{\mu_c}}}
= \sum_N \frac{\mu^N}{N!}\,\,\frac{C (2N-1)!!}{2^N \mu_c^N} ,
\end{equation}
and comparing with:
\begin{equation}
\partial_\mu^3\hat{\mathcal Z}(\mu,L)=\sum_{N}\frac{\mu^N}{N!}\mathcal{Z}_{N+3}(L) ,
\end{equation}
we find the large $N$ behavior of the volume
\begin{equation}
\mathcal Z_{N+3}(L) \sim C\,\frac{(2N-1)!!}{2^N \mu_c^N}
= C\,\frac{(2N-1)!!\,L^{2N}}{2^N (\mu_c L^2)^N}
\qquad , \qquad
\mu_c L^2 = 1.902...
\end{equation}

\subsection{One-point function -- Saddle point method}

We want to study the large $N$ limit of the one-point function:
\begin{align}
f_N\left(L,\frac{L_1}{L}\right)&\overset{\mathrm{def}}{=}\mathcal{Z}_{N+3}(L,L_1)\cr 
%&=\frac{1}{2^{N+1}}\int_{\mathcal{M}_{0,N+3}}e^{\frac{L^2}{2}\sum_{i=2}^{N+3}\psi_i+\frac{L_1^2}{2}\psi_1}& \\
&= \dots\\
&= \frac{N!L^{2N}}{2}\underset{z\to 0}{\Res}\frac{dz}{z}I_0^2(z)e^{N\left(\ln I_0(z)-2\ln z + \frac{1}{N}\ln I_0\left(\frac{L_1}{L}z\right)\right)}\\
\end{align}
The detail of the computation has been transferred to appendix \ref{appOnePoint} for readability, as the calculus is close to the one for the volume.
Let us define
\begin{equation}
S_N(z)=\ln I_0(z)-2\ln z + \frac{1}{N}\ln I_0\left(\frac{L_1}{L}z\right)
\end{equation}
$S_N$ is an even function. 
In the large $N$ limit, we use the saddle point approximation to compute the residue, hence we have to find the saddle point of $S_N$.
First, let us compute its derivatives.
\begin{align}
\frac{\partial}{\partial y} S_N(x+i y)&=&i\left[\frac{I_1(x+i y)}{I_0(x+i y)}-\frac{2}{x+i y}+\frac{1}{N}\frac{L_1}{L}\frac{I_1\left(\frac{L_1}{L}(x+i y)\right)}{I_0\left(\frac{L_1}{L}(x+i y)\right)}\right]\\
\frac{\partial^2}{\partial y^2} S_N(x+i y)&=&-1+\frac{I_1(x+i y)}{(x+i y)I_0(x+i y)}+\frac{I_1^2(x+i y)}{I_0^2(x+i y)}-\frac{2}{(x+i y)^2}-\\
& &\frac{1}{N}\left(\frac{L_1}{L}\right)^2\left(1-\frac{I_1\left(\frac{L_1}{L}(x+i y)\right)}{\frac{L_1}{L}(x+i y)I_0\left(\frac{L_1}{L}(x+i y)\right)}-\frac{I_1^2\left(\frac{L_1}{L}(x+i y)\right)}{I_0^2\left(\frac{L_1}{L}(x+i y)\right)}\right) \\
\end{align}
We distinguish three regimes for the behaviour of $L_1$ at large $N$. For each regime, we may compute the saddle points and carry out the residue.

%\subsubsection{Regime 1: $\frac{1}{N}\frac{L_1}{L}\underset{N\to\infty}{\to}0$}
\subsubsection{Regime 1: $L_1/(NL)\to 0$ when $N\to\infty$}

In this regime, the term $\frac1 N \ln{I_0\left(\frac{L_1}{L} z\right)}$ is negligible, the saddle point is the saddle point of $\ln I_0(z)-2\ln z$, it is independent of $L_1/L$, and it is worth $z=\pm u_c$.
This gives
\begin{align}
f_N\left(L,\frac{L_1}{L}\right)
&\sim 
I_0\left(\frac{L_1}{L} u_c\right) \, \frac{I_0(u_c)^{N+2}}{u_c^{2N}}\, \frac{\sqrt{2\pi}}{\sqrt{u_c^2-4}} \, N!L^{2N} \\
& \propto I_0\left(\frac{L_1}{L} u_c\right)
\end{align}
It thus behaves like Bessel function $I_0(u_c L_1/L)$.

\subsubsection{Regime 2: ${L_1}/{NL}\to l$ when $N\to\infty$}

We use the asymptotics:
\begin{equation}
I_0(x)\underset{x\to\infty}{=}\frac{e^x}{\sqrt{2\pi x}}(1+O\left(\frac{1}{x}\right))
\end{equation}
which gives: 
\begin{equation}
S_N(z)=\ln\left(\frac{I_0(z)}{z^2}\right)+ l z+O\left(\frac{\ln N}{N}\right)
\end{equation}
By the same argument as in the first regime, there are two saddle points $x_0(l)$, $x_1(l)=-x_0(l)$ situated on the real axis. Again, let $x_0$ be the positive one.
The equation $S'_N(x_0)=0$ gives:
\begin{equation}
\label{lx0}
x_0 I_1(x_0)-\left(2-l x_0\right)I_0(x_0)=0
\end{equation}
At the point $x_0(l)$:
\begin{align}
S_N(x_0)&=\ln I_0(x_0)-2\ln x_0+l x_0+o(1)\\
\frac{\partial^2}{\partial y^2}S_N(x_0)&=-\left(1+\frac{4l}{x_0}-\frac{4}{x_0^2}-l^2\right)+o(1)\\
&=-1+\left(l-\frac{2}{x_0}\right)^2+o(1)\\
&= O(1)
\end{align}

We then have:
\begin{equation}
f_N\left(L,\frac{L_1}{L}\right)=N! L^{2N} \frac{I_0(x_0(l))^{N+2}}{x_0(l)^{2N}}\frac{e^{Nlx_0(l)}}{\sqrt{2\pi Nlx_0(l)}}\frac{\sqrt{2\pi}}{\sqrt{(l^2+1)x_0(l)^2+5x_0(l)-4}}
\end{equation}
Of course, the factors $2\pi$ simplify, but in this form, we see that $f_N$ in the second regime is matching the one of the first regime. Indeed, as $\frac{L_1}{L}\sim N l$, we have:
\begin{equation}
I_0\left(\frac{L_1}{L}x_0(l)\right)\sim \frac{e^{Nlx_0(l)}}{\sqrt{2\pi Nlx_0(l)}}
\end{equation}
What is more, if $l=0$, the last fraction is equal to $ \frac{\sqrt{2\pi}}{\sqrt{u_c^2-4}}$. So we recover the first regime in this limit.

%There is one pair $(l^c,x_0^c)$ which satisfies:
%\begin{align}
%\ln I_0(x_0^c)-2\ln x_0^c + 2 l^c x_0^c=0\\
%x_0^c I_1(x_0^c)-(2-l^c x_0^c)I_0(x_0^c)=0
%\end{align}
%Then, there are two cases:
%\begin{itemize}
%\item If $l< l^c$, $S_N(x_0)<0$ and then, applying the saddle point approximation, we get the same limit for the residue as in the first regime:
%    \begin{equation}
%    \underset{z\to 0}{\Res}\frac{dz}{z}I_0^2(z)e^{N\left(\ln I_0(z)-2\ln z + \frac{1}{N}\ln I_0\left(\frac{L_1}{L}z\right)\right)}\underset{N\to \infty}{ \sim} C_1 \frac{C_2^N}{\sqrt{N}}
%    \end{equation}
%    with $C_2<1$
%\item If $l>l^c$, $S_N(x_0)>0$, so 
%    \begin{equation}
%    \underset{z\to 0}{\Res}\frac{dz}{z}I_0^2(z)e^{N\left(\ln I_0(z)-2\ln z + \frac{1}{N}\ln I_0\left(\frac{L_1}{L}z\right)\right)}\underset{N\to \infty}{ \sim} C_1 \frac{C_2^N}{\sqrt{N}}
%    \end{equation}
%    with $C_2>1$
%\end{itemize}

%$C_2$ depends on $l$. From equation \ref{lx0} we get 
%\begin{equation}
%l=\frac{2I_0(x_0)-x_0 I_1(x_0)}{x_0 I_0(x_0)}
%\end{equation}
%So we can write $x_0$ as a function of $l$. Then
%\begin{equation}
%C_2(l)=e^{\ln I_0(x_0(l))-2\ln x_0(l)+l x_0(l)}
%\end{equation}

%\subsubsection{Regime 3:  $\frac{1}{N}\frac{L_1}{L}\underset{N\to\infty}{\to}\infty$ }
\subsubsection{Regime 3:  $L_1/(NL)\to\infty$ when $N\to\infty$}

We note $l=\frac{1}{N}\frac{L_1}{L}$, so in this regime, $l\gg1$.
We can show that in this regime, we have necessarily, for the saddle point $x_0$:
\begin{align}
x_0\underset{N\to\infty}{\to}0\\
\frac{L_1}{L}x_0 \underset{N\to \infty}{\to} +\infty
\end{align}
We can then expand $x_0$ as a series of $N^\alpha \left(\frac{L_1}{L}\right)^\beta$. We find:
\begin{equation}
x_0(l)=\frac{2}{l}+\frac{2}{5}\frac{1}{N l}+O\left(\frac{1}{N^2 l}\right)
\end{equation}
We then get:
\begin{equation}
\frac{\partial^2 S}{\partial y^2}(x_0(l))=-\frac{l^2}{2}(1+O\left(\frac{1}{N}\right))
\end{equation}
and
\begin{equation}
S(x_0(l))=\ln I_0\left(\frac{2}{l}\right)-2\ln \frac{2}{l}+\frac{1}{N}\ln I_0(2N)
\end{equation}
In the end, we obtain:
%\begin{equation}
%\underset{z\to 0}{\Res}\frac{dz}{z}I_0^2(z)e^{N\left(\ln I_0(z)-2\ln z + \frac{1}{N}\ln I_0\left(\frac{L_1}{L}z\right)\right)}\underset{N\to \infty}{ \sim} C \frac{1}{N}\frac{L_1}{L}\frac{e^{2N\ln\left(\frac{1}{N}\frac{L_1}{L}\right)}}{\sqrt{N}}
%\end{equation}
%with $C$ a constant.
\begin{equation}
f_N\left(L,\frac{L_1}{L}\right)=N! L^{2N} I_0\left(\frac{2}{l}\right)^{N+2}\left(\frac{l}{2}\right)^{2N} \sqrt{\pi}I_0(2N)
\end{equation}

%\vfill\eject

\subsection{Correlation functions from the Spectral Curve}
In \cite{Eynard2011} it was shown that the Eynard-Orantin invariants of the spectral curve, are generating functions of intersection numbers
\begin{equation}
2^{2-2g-n} \sum_{d_1,\dots,d_n}
\prod_{i=1}^n \frac{(2d_i-1)!! }{z_i^{2d_i+1}} 
\left\langle
e^{\frac12 \sum_k (2k-1)!! \tilde t_{2k+1} \tau_k}
\prod_{i=1}^n \tau_{d_i}
\right\rangle_g
= F_{g,n}(\mathcal S;z_1,\dots,z_n)
\end{equation}
This is true in particular for $g=0$.
We have
\begin{equation}
F_{0,n}(\mathcal S;z_1,\dots,z_n) 
= \mathcal F_{n}(\mu,L;z_1,\dots,z_n) .
\end{equation}

We recall here a theorem that we will use for asymptotics of $n$-point functions with $n\geq3$. It is the theorem of section 8 in \cite{EynardOrantin2007}, proven by Eynard and Orantin. It states that, if $2-2g-n<0$ and the spectral index $\mu$ is close to its critical value $\mu_c$, the Eynard-Orantin invariants diverge as:
\begin{eqnarray}
F_{g,n}(\mathcal S;z_1,\dots,z_n)
&\underset{\mu\to\mu_c}{\sim} & (u_c-u)^{(2-2g-n)(1+\frac32)} \left(\frac{u_c}{L^2}\frac{u_c^2-4}{6L\sqrt{u_c}} \right)^{2-2g-n}\times\cr
& &F_{g,n}({\mathcal S}_{(3,2)};\xi_1,\dots,\xi_n) \cr
&\underset{\mu\to\mu_c}{\sim}& (u_c-u)^{(2-2g-n)\frac52} \left(\frac{u_c}{L^2}\,\frac{u_c^2-4}{6L\sqrt{u_c}} \right)^{2-2g-n}
F_{g,n}({\mathcal S}_{(3,2)};\xi_1,\dots,\xi_n) \cr
&\underset{\mu\to\mu_c}{\sim}& (1-\mu/\mu_c)^{(2-2g-n)\frac54} \left( \left(\frac{2u_c^2}{u_c^2-4}\right)^{\frac54}\,  \frac{u_c}{L^3}\,\frac{u_c^2-4}{6\sqrt{u_c}} \right)^{2-2g-n}\times\cr
& &F_{g,n}({\mathcal S}_{(3,2)};\xi_1,\dots,\xi_n) \cr
\end{eqnarray}

Again the exponent $5/4=(p+q)/(p+q-1)$ is the KPZ exponent \cite{KPZ} for the $(p,q)$ minimal model coupled to gravity, and here $(p,q)=(3,2)$.

\subsubsection{The one-point function}
Here, we compute the same quantity as in the previous section, from the spectral curve. The asset of such a method is that it generalizes easily to higher correlation functions.
The quantity $f_N\left(\frac{L_1}{L}\right)$ is encoded in the following generating function:
\begin{equation}
H(\mu,L,L_1)=\sum_{N=0}^{\infty}\frac{\mu^{N+3}}{(N+3)!}f_N\left(L,\frac{L_1}{L}\right)
\end{equation}
In terms of the Chern classes, we can express $f_N $ as:
\begin{equation}
f_N\left(L,\frac{L_1}{L}\right)=\frac{1}{2^{N+1}}\int_{\mathcal{M}_{0,+3}}e^{\frac{L^2}{2}\sum_{i=2}^{N+3}\psi_i+\frac{L_1^2}{2}\psi_1}
\end{equation}
It is related to the differential $\mathcal{W}_1(\mu,L;z_1)$ by:
\begin{equation}
H(\mu,L,L_1)=\sum_{d_1=0}^{\infty}L_1^{2d_1} \underset{z_1 \to \infty}{\mathrm{Res}}\frac{z_1^{2d_1+1}}{(2d_1+1)!}\mathcal{W}_1(\mu,L;z_1)
\end{equation}
Setting $x_1=z_1^2$, we can rewrite it in the following way:
\begin{equation}
H(\mu,L,L_1)=\sum_{d_1=0}^{\infty}L_1^{2d_1} \underset{x_1 \to \infty}{\mathrm{Res}}\frac{x_1^{d_1+\frac{1}{2}}}{(2d_1+1)!}\mathcal{W}_1(\mu,L;x_1)
\end{equation}
The differential $\mathcal{W}_1=ydx$ is the one point function of our model with times $t_k$ (see section 3), in which $t_1=\mu\neq0$. To get rid of $t_1$, we have renormalized the times into $\check{t}_i$, and then the one point function is given by $y(z)dx(z)$ (the spectral curve being given by $(x(z),y(z))$). In our model, $x=z^2+\frac{u^2}{L^2}$. In the end, we have to compute the following:
\begin{equation}
H(\mu,L,L_1)=\sum_{d_1=0}^{\infty}L_1^{2d_1} \underset{z \to \infty}{\mathrm{Res}}\frac{(z^2+\frac{u^2}{L^2})^{d_1+\frac{1}{2}}}{(2d_1+1)!}y(z)dx(z)
\end{equation}
The function $y(z)$ is an entire function of $z$, with only poles at infinity. The differential $dx(z)=2zdz$ also has no pole except at $\infty$. We may then deform the contour of integration of the residue. The quantities $(z^2+\frac{u^2}{L^2})^{d_1+\frac{1}{2}}$ have branch cuts. We choose as branch cut for $x\mapsto\sqrt{x}$ the half line $i\mathbb{R}_{-}$. Then $(z^2+\frac{u^2}{L^2})^{d_1+\frac{1}{2}}$ has one cut, along the segment $[-i\frac{u}{L},+i\frac{u}{L}]$. \\
Let us deform the contour of the residue around this segment and call it $\mathcal{C}$. The following residue is null:
\begin{equation}
\frac{1}{2i\pi}\oint_{\mathcal{C}}(z^2+\frac{u^2}{L^2})^{d_1}y(z)dx(z)=0
\end{equation}
for any integer $ d_1$, as it does not enclose any pole or cut. We can then add to the sum the following sum without changing the function $H$:
\begin{equation}
\sum_{d_1=0}^{\infty}\frac{L_1^{2d_1-1}}{(2d_1)!}\frac{1}{2i\pi}\oint_{\mathcal{C}}(z^2+\frac{u^2}{L^2})^{d_1}y(z)dx(z)
\end{equation}
So:
\begin{equation}
H(\mu,L,L_1)=-\frac{1}{L_1}\sum_{d_1=0}^{\infty}\frac{L_1^{d_1}}{d_1 !}\frac{1}{2i\pi}\oint_{\mathcal{C}}\sqrt{z^2+\frac{u^2}{L^2}}^{d_1}y(z)dx(z).
\end{equation}
We can exchange $\sum$ and $\oint$, and it remains to compute:
\begin{equation}
H(\mu,L,L_1)=-\frac{1}{L_1}\frac{1}{2i\pi}\oint_{\mathcal{C}}e^{\sqrt{z^2+\frac{u^2}{L^2}}L_1}y(z)dx(z)
\end{equation}
Let us decompose the contour $\mathcal{C}$ into  $\mathcal{C}^+ = [-i\frac{u}{L}+\epsilon,+i\frac{u}{L}+\epsilon]$ and  $\mathcal{C}^- = [-i\frac{u}{L}-\epsilon,+i\frac{u}{L}-\epsilon]$, with $\epsilon>0$ and small. Then:
\begin{eqnarray}
H(\mu,L,L_1)&=&-\frac{1}{L_1}\frac{1}{2i\pi}\left(\int_{\mathcal{C}^+}-\int_{\mathcal{C}^-}\right)e^{\sqrt{z^2+\frac{u^2}{L^2}}L_1}y(z)dx(z)\\
&=&-\frac{1}{L_1}\frac{1}{2i\pi}\int_{[-i\frac{u}{L},+i\frac{u}{L}]}2\sinh{\left(\sqrt{z^2+\frac{u^2}{L^2}}L_1\right)}y(z)dx(z)
\end{eqnarray}
(in the last line, the cut of $\sqrt{\,\,}$ is $\mathbb{R}_-$).\\
\begin{equation}
H(\mu,L,L_1)=-\frac{1}{L_1}\frac{1}{2i\pi}\int_{-i\frac{u}{L}}^{+i\frac{u}{L}}2\sinh{\left(\sqrt{z^2+\frac{u^2}{L^2}}L_1\right)}\left[z^2-\frac{\mu}{2}\sum_{k=1}^{\infty}\frac{I_k(u)L^{2k}}{(2k+1)!!u^k}z^{2k+2}\right]dz
\end{equation}
This last integral is explicitly computable. In order to do that, we use the result:
\begin{equation}
\int_{0}^{a}t \sqrt{a^2-t^2}^{2k+1}\sinh{t}\,dt=\frac{\pi}{2}a^{k+2}(2k+1)!! I_{k+2}(a)
\end{equation}
In the end, we obtain:
\begin{equation}
H(\mu,L,L_1)=\frac{1}{L_1}\frac{u^2}{L^2}\left[I_2(u\frac{L_1}{L})-\frac{\mu}{2}\sum_{k=1}^{\infty}\left(-\frac{L}{L_1}\right)^k I_k(u)I_{k+2}(u\frac{L_1}{L})(2k+1)\right]
\end{equation}
As $\mu\to\mu_c$, $H$ remains finite (this is not true anymore for $n-$point functions with $n\geq3$) and has a non null limit, but has a singular term in $(\mu_c-\mu)^{\frac{1}{2}}$ in its expansion near $\mu_c$.

\subsubsection{n-point functions for $n\geq$ 3.}
\begin{figure}
\centering
\includegraphics[width=5cm]{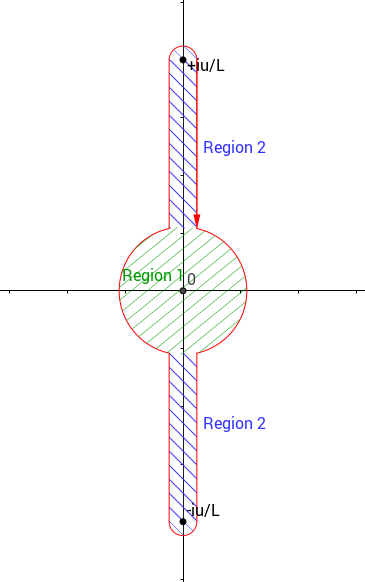}
\caption{The contour of integration of $\underset{\infty}{\mathrm{Res}}$ is deformed and encloses a cut.}
\label{contour_cut}
\end{figure}

For $n\geq3$, we have:
\begin{equation}
2^{2-n}\prod_{i=1}^{n}(2d_i-1)!! \langle e^{\frac{\mu}{2}\sum_d \frac{L^{2d}}{2^d d!}\tau_d}\prod_{i=1}^{n}\tau_{d_i}\rangle_{0}=\underset{z_i\to\infty}{\mathrm{Res}}\prod_{i=1}^{n}z_i^{2d_i}dz_i F_{0,n}(\mathcal{S};z_1,\dots,z_n).
\end{equation}
The quantity we are interested in is $\hat{\mathcal{Z}}_n(\mu,L;L_1,\dots,L_n)$ and we want to recover it from the previous expression. The left hand side is equal to:
\begin{equation}
\mathrm{l.h.s.}=2^{-\sum_i d_i} L^{-2D}\prod_{i=1}^{n}(2d_i -1)!! \frac{\partial^{n-3}}{\partial\mu^{n-3}}[\mu^n\mathcal{U}_n(\mu,L,d_1,\dots,d_n)]
\end{equation}
Hence, in order to recover $ \hat{\mathcal{Z}_n} $, we have to carry out the following sum:
\begin{eqnarray}
\hat{\mathcal{Z}_n}(\mu,L;L_1,\dots,L_n)&=&\sum_{d_1\dots d_n}\prod_{i=1}^{n}\frac{L_i^{2d_i}}{2^{d_i}(2d_i-1)!! d_i!}\times \mathrm{l.h.s.}\\
&=& \sum_{d_1\dots d_n}\prod_{i=1}^{n}\frac{L_i^{2d_i}}{(2d_i)!}\times \underset{z_i\to\infty}{\mathrm{Res}}\prod_{i=1}^{n}z_i^{2d_i}dz_i F_{0,n}(\mathcal{S};z_i).
\end{eqnarray}
The functions $F_{0,n}(\mathcal{S},z_i)$ are entire functions of the $\frac{1}{z_i}$. By the following change of coordinates : $z_i=\sqrt{\tilde{z}_i^2+\frac{u^2}{L^2}}$, the functions $F_{0,n}(\mathcal{S};\tilde{z}_i)$ are then polynomials in the $\frac{1}{\tilde{z}_i} $. This change of variable is similar to the one done for the one-point function of the previous section. Indeed, the variables $\tilde{z}_i^2+\frac{u^2}{L^2}$ are the $ x_i $, living on the spectral curve $ \mathcal{S}$, and are then more 'canonical' than the $z_i$. So we have:
\begin{equation}
\hat{\mathcal{Z}}_n(\mu,L;L_i)=\sum_{d_1\dots d_n}\prod_{i=1}^{n}\frac{L_i^{2d_i}}{(2d_i)!}\underset{\tilde{z}_i\to\infty}{\mathrm{Res}}\prod_{i=1}^{n}\left(\tilde{z}_i^2+\frac{u^2}{L^2}\right)^{d_i-\frac{1}{2}}2\tilde{z}_i d\tilde{z}_i F_{0,n}(\mathcal{S};\tilde{z_i})
\end{equation}
The residue is taken at infinity, and we want to deform its contour of integration. The function $F_{0,n}(\mathcal{S};\tilde{z}_i)$ has poles only around $\tilde{z}_i=0$ ; the term $\left(\tilde{z}_i^2+\frac{u^2}{L^2}\right)^{d_i-\frac{1}{2}} $ has a cut on the segment $\left[-i\frac{u}{L};+i\frac{u}{L}\right]$ (the branch cut for $\sqrt{\,}$ is $-i\mathbb{R}$). Hence, we can deform the contour of integration into the one described in figure \ref{contour_cut}. \\

Now it is possible to exchange $\sum$ and $\mathrm{Res}$, because the contour is no longer at infinity. This gives:
\begin{eqnarray}
\hat{\mathcal{Z}}_n(\mu,L;L_i)&=&\sum_{d_1\dots d_n}\underset{\tilde{z}_i\to\infty}{\mathrm{Res}} \prod_{i=1}^{n}\frac{L_i^{d_i}}{d_i !}\sqrt{\tilde{z}_i^2+\frac{u^2}{L^2}}^{d_i}\frac{2\tilde{z}_i F_{0,n}(\mathcal{S};\tilde{z}_i)}{\sqrt{\tilde{z}_i^2+\frac{u^2}{L^2}}}d\tilde{z}_i\\
&=&2^n \underset{\tilde{z}_i\to\infty}{\mathrm{Res}}\prod_{i=1}^{n}e^{L_i\sqrt{\tilde{z}_i^2+\frac{u^2}{L^2}}}\frac{\tilde{z}_i F_{0,n}(\mathcal{S};\tilde{z}_i)}{\sqrt{\tilde{z}_i^2+\frac{u^2}{L^2}}}d\tilde{z}_i
\end{eqnarray}

\begin{figure}
\centering
\includegraphics[width=4cm]{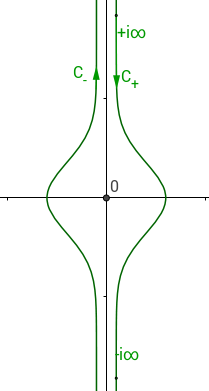}
\caption{Contour of the region 1 in the variable $\xi_i$.}
\label{limiting_contour}
\end{figure}

We want the asymptotic behaviour ($N\to\infty$) of the Strebel Graph volumes with $n$ marked faces, which corresponds to look at the limit $\mu\to\mu_c$ in the function $\hat{\mathcal{Z}}_n$. In that limit, the contour can be divided into two regions (see figure \ref{contour_cut}) : region 1 corresponds to the parts of the contour which are close to the pole ($\tilde{z}_i=0$) of $F_{0,n}$,  region 2 corresponds to the rest of the contour.\\
The contour integral over region 2 remains finite (of order 1) when $\mu\to\mu_c$. As we may see in the following, on the contrary, the integral over region 1 diverges as $\mu\to\mu_c$. \\
Region 1 is the part of the integral close to 0. Let us define:
\begin{equation}
\tilde{z}_i=-\sqrt{u_c-u}\frac{\sqrt{u}}{L}\xi_i
\end{equation}
From the theorem of section 8 of \cite{EynardOrantin2007}, we have, in the limit $\mu\to\mu_c$ (and hence $\tilde{z}_i\to0$):
\begin{eqnarray}
F_{0,n}(\mathcal{S};\tilde{z}_i)\underset{\mu\to\mu_c}{\sim}(u_c-u)^{\frac{5}{2}(2-n)}\left(\frac{u_c}{L^2}\frac{u_c^2-4}{6L\sqrt{u_c}}\right)^{2-n}F_{0,n}(\mathcal{S}_{(3,2)};\xi_i)
\end{eqnarray}
where $F_{0,n}(\mathcal{S}_{(3,2)};\xi_i)$ are the invariants of the model (3,2). We see that this quantity behaves like $(u_c-u)^{\frac{5}{2}(2-n)}$, so, as $n\geq3$, it is divergent as $ \mu\to\mu_c$.\\
In order to get the dominant order in the large N limit, we may then focus our attention on the region 1. In the variables $ \xi_i$, we have to carry out the integration over the contours $\mathcal{C}_+,\,\mathcal{C}_-$ (see figure \ref{limiting_contour}). $\mathcal{C_+}$ is going from $+i\infty$ to $-i\infty$, with $Re(\xi_i)>0$ ; $\mathcal{C_-}$ is going from $-i\infty$ to $+i\infty$, with $Re(\xi_i)<0$.\\

We look at the expansion in $\mu_c-\mu$, so we reexpress the square roots as:
\begin{eqnarray}
\sqrt{\tilde{z}_i^2+\frac{u^2}{L^2}}&=&\sqrt{\frac{u_c^2}{L^2}+(u_c-u)\frac{u_c}{L^2}(\xi_i^2-2)+O((u_c-u)^2)}\\
&=&\frac{u_c}{L}\sqrt{1+\frac{u_c-u}{u_c}(\xi_i^2-2)+O((u_c-u)^2)}\\
&=&\frac{u_c}{L}+\frac{1}{2}\frac{u_c-u}{L}(\xi_i^2-2)+ O((u_c-u)^2)
\end{eqnarray}
We are also looking at a regime where $\frac{L_i}{L}\to\infty$ as $\mu\to\mu_c$. From the previous expansion, we see that the argument of the exponential contains $ \frac{L_i}{L}(u_c-u)(\xi_i^2-2)$, which, at $\xi_i$ fixed, remains of order 1 if $\frac{L_i}{L}\sim(u_c-u)^{-1}\sim(\mu_c-\mu)^{-\frac{1}{2}}$. This corresponds to a regime where $ \frac{L_i}{L}\sim \sqrt{N}$.\\

The function to integrate is odd, but $\mathcal{C}_+$ and $\mathcal{C}_-$ have opposite orientations, so we can restrict to the integration over $\mathcal{C}_+$:
\begin{eqnarray}
\hat{\mathcal{Z}}_n(\mu,L;L_i)&\underset{\mu\to\mu_c}{\sim}& 2^n(u_c-u)^{\frac{5}{2}(2-n)}\left(\frac{u_c}{L^2}\frac{u_c^2-4}{6L\sqrt{u_c}}\right)^{2-n}\frac{2}{2i\pi}\int_{+i\infty}^{-i\infty}d\xi_1\dots\frac{2}{2i\pi}\int_{+i\infty}^{-i\infty}d\xi_n\cr
& &\prod_{i=1}^{n}e^{u_c\frac{L_i}{L}+\frac{1}{2}\frac{L_i}{L}(u_c-u)(\xi_i^2-2)}(u_c-u)\frac{u}{L^2}\frac{\xi_i}{\frac{u_c}{L}+\frac{1}{2}\frac{u_c-u}{L}(\xi_i^2-2)}\times\cr
& & F_{0,n}(\mathcal{S}_{(3,2)};\xi_i)(1+ O((u_c-u)^2))\cr
&\underset{\mu\to\mu_c}{\sim}& \frac{2^{2n}}{(2i\pi)^n}(u_c-u)^{5-\frac{3}{2}n}\left(\frac{u_c^2-4}{6L\sqrt{u_c}}\right)^{2-n}\left(\int_{+i\infty}^{-i\infty}d\xi_1\dots\int_{+i\infty}^{-i\infty}d\xi_n\right)\cr
& & \prod_{i=1}^{n}e^{u_c\frac{L_i}{L}+\frac{1}{2}\frac{L_i}{L}(u_c-u)(\xi_i^2-2)}\frac{\xi_i}{\frac{u_c}{L}} F_{0,n}(\mathcal{S}_{(3,2)};\xi_i)(1+O(u_c-u))\cr
\end{eqnarray}
We carry out the change of variable $\xi_i=i\zeta_i$, and in the end:
\begin{eqnarray}
\hat{\mathcal{Z}}_n(\mu,L;L_i)&\underset{\mu\to\mu_c}{\sim}& \frac{2^{n}}{i^n\pi^n}(u_c-u)^{5-\frac{3}{2}n}\left(\frac{u_c^2-4}{6L\sqrt{u_c}}\right)^{2-n}\frac{L^n}{u_c^n}\int_{+\infty}^{-\infty}id\zeta_1\dots\int_{+\infty}^{-\infty}id\zeta_n\cr
& & \prod_{i=1}^{n}e^{u\frac{L_i}{L}}\prod_{j=1}^{n}e^{-\frac{1}{2}\frac{L_j}{L}(u_c-u)\zeta_j^2}i\zeta_j F_{0,n}(\mathcal{S}_{(3,2)};i\zeta_j)(1+O(u_c-u))\cr
&\underset{\mu\to\mu_c}{\sim}& \frac{2^{n}}{i^n\pi^n}(u_c-u)^{5-\frac{3}{2}n}\left(\frac{u_c^2-4}{6L\sqrt{u_c}}\right)^{2-n}\frac{L^n}{u_c^n}\int_{-\infty}^{+\infty}d\zeta_1\dots\int_{-\infty}^{+\infty}d\zeta_n\cr
& & \prod_{i=1}^{n}e^{u\frac{L_i}{L}}\prod_{j=1}^{n}e^{-\frac{1}{2}\frac{L_j}{L}(u_c-u)\zeta_j^2}\zeta_j F_{0,n}(\mathcal{S}_{(3,2)};i\zeta_j)(1+O(u_c-u)).\cr
\end{eqnarray}
The function $F_{0,n}$ being a polynomial in $\frac{1}{\zeta_j}$, the result, as we shall see with $\hat{\mathcal{Z}}_3$ and $\hat{\mathcal{Z}}_4$, is expressible in terms of Gamma functions.
\subsubsection{3-point function and 4-point function in the large N limit}
In the (3,2) model, we have:
\begin{eqnarray}
F_{0,3}(\mathcal{S}_{(3,2)};\xi_1,\xi_2,\xi_3)&=&\frac{1}{6\xi_1\xi_2\xi_3}\\
F_{0,4}(\mathcal{S}_{(3,2)};\xi_1,\xi_2,\xi_3,\xi_4)&=&-\frac{1}{36\xi_1\xi_2\xi_3\xi_4}\left[1+\frac{1}{\xi_1^2}+\frac{1}{\xi_2^2}+\frac{1}{\xi_3^2}+\frac{1}{\xi_4^2}\right]
\end{eqnarray}
Applying the result of the previous section, we obtain:
\begin{equation}
\hat{\mathcal{Z}}_3(\mu,L;L_i)\underset{\mu\to\mu_c}{\sim}8\left(\frac{2}{\pi}\right)^{\frac{3}{2}}\frac{L^4}{u_c^{\frac{5}{2}}(u_c^2-4)}(u_c-u)^{\frac{1}{2}}\prod_{i=1}^{3}\frac{e^{u_c\frac{L_i}{L}}}{\sqrt{\frac{L_i}{L}(u_c-u)}}
\end{equation}
It may seem that this quantity is not divergent, but remember that, in the exponentials, we have $\frac{L_i}{L}\sim(u_c-u)^{-1}$.\\
For the 4-point function:
\begin{equation}
\hat{\mathcal{Z}}_4(\mu,L;L_i)\underset{\mu\to\mu_c}{\sim} \frac{64}{\pi^2}\frac{L^6}{u_c^3}\frac{1}{(u_c^2-4)^2}(u_c-u)^{-1}\left(1+\sum_{i=1}^{4}\frac{L_i}{L}(u_c-u)\right)\prod_{i=1}^{4}\frac{e^{u_c\frac{L_i}{L}}}{\sqrt{\frac{L_i}{L}(u_c-u)}}
\end{equation}
Here, the divergence is clear. We want to underline that the terms $ \sum_{i}\frac{L_i}{L}(u_c-u)$ are not subdominant, but of order 1, so we have to take them into account.

\section{Conclusion}

We showed that in the large $N$ limit (number of vertices) the expectation values (over the set of Strebel graphs with fixed perimetres and the Kontsevich measure) of all the topological observables (algebraic combinations of the Chern classes) converge to the corresponding Liouville CFT amplitudes at central charge $c=0$, i.e. quantum gravity, equivalent to the $(3,2)$ minimal model.
In particular we recovered the KPZ exponents.
Moreover, we found explicit expressions of all those amplitudes at finite $N$, as well as their explicit asymptotics at $N\to\infty$, in various regimes.

Our method could be easily generalized to genus one graphs, since all intersection numbers of genus one are explicitly known, and generating functions are also Bessel functions.
The spectral curve methods works for all genus and shows that the continuum limit tends to the (3,2) minimal model's result for all genus.
It would be interesting to extend results in higher genus cases.

\section*{Aknowledgements}

BE was supported by the ERC Starting Grant no. 335739 ``Quantum fields and knot homologies'' funded by the European Research Council under the European Union's Seventh Framework Programme.  
BE is also partly supported by the ANR grant Quantact : ANR-16-CE40-0017.
We thank P. di Francesco for his interest and useful conversations.

\appendix
\section{Metric associated to a Strebel graph}\label{appStrebel}
To every Strebel graph is associated a unique metric. \\ 
As we mentionned in the first part, and as was proven by Strebel, Penner, Zaguier and Kontsevich, the set of Strebel graphs $\oplus_{G\in\mathcal{G}_{0,N+3}}\mathbb{R}_+^{\mathcal{E}(G)}$ is in bijection with $\tilde{\mathcal{M}}_{0,N+3}$.  
A point in $\tilde{\mathcal{M}}_{0,N+3}$ is a set $\{z_1,\dots,z_{N+3}\} $ of distinct complex numbers, with $\{z_1,z_2,z_3\}=\{0,1,\infty\}$, along with $N+3$ positive perimeters $L_1,\dots,L_{N+3}$. 
Let us define a meromorphic \textbf{quadratic differential} $\Omega(z)=f(z)dz^2$ with $f(z)$ a rational function having $N+2$ double poles in $z_i,\, i=1,2,4,\dots,N+3$, and behaving like $O(1/z^2)$ near $z_3=\infty$. We hence impose the following behaviour to $f(z)$:
\begin{align}\label{behaveDiff}
i\neq 3 \ \qquad &\to \quad
f(z)\underset{z\to z_i}{=}\frac{-L_i^2}{(z-z_i)^2}(1+O(z-z_i)) \cr
i= 3 \ \qquad &\to \quad
f(z)\underset{z\to \infty}{=}\frac{-L_3^2}{z^2}(1+O(1/z)).
\end{align}
The generic $f$ is:
\begin{align}
f(z) = \frac{-1}{\prod_{i\neq 3} (z-z_i)} \ \left( \sum_{i\neq 2} \frac{L_i^2 \ \prod_{j\neq i,3} (z_i-z_j)}{z-z_i} + L_3^2 z^N + p_{N-1}(z)\right) \cr
\text{where} \ \ p_{N-1}=\text{polynomial of degree }\leq N-1.
\end{align}

The level lines --called horizontal trajectories-- of this differential are the lines where 
\begin{equation}
\operatorname{Im}\left(\int^x \sqrt{f(z)}dz\right)=\mathrm{constant}.
\end{equation}
Almost all the closed level lines are topological circles surrounding one or several double poles. The other closed trajectories --called critical trajectories-- form a graph. 
The Strebel theorem (\cite{Strebel1984}) states that, to a point in $\tilde{\mathcal{M}}_{0,N+3}$ corresponds a unique quadratic differential $\Omega$ --called \textbf{Strebel differential}-- having the same behaviour as defined in equation \ref{behaveDiff}, and such that the graph formed by the critical trajectories is cellular on the surface: each face is a topological disc and contains exactly one double pole.\\
The metric associated to a Strebel graph is then the flat metric :
\begin{equation}
g_{zz}=g_{\bar{z}\bar{z}}=0,\qquad g_{z\bar{z}}=\overline{g_{\bar{z} z} }=\frac{1}{2} f(z) .
\end{equation}
The lengths $\ell_e$ of the Strebel graph edges are the lengths measured with the metric $g$ along the level lines joining the vertices of the graph.\\
The metric is flat on $\mathbb{C}\backslash(\{z_1,\dots,z_{N+3}\}\cup\{\mathrm{zeros\,of\,}\Omega\})$.
Its curvature is a distribution localized at the $N+3$ poles $z_i$s (curvature $2\pi$) and the $2N+2$ vertices (curvature =$-\pi$), so that the total curvature is $4\pi=2\pi\chi$ with the Euler characteristic $\chi=2-2g=2$ for a genus zero surface.

The Riemann surface with $N+3$ punctures can be realized by  $N+3$ semi--infinite cylinders of respective perimeters $L_1,\dots,L_{N+3}$  glued to the graph along their bases.

\section{Explicit computation of the one point function}\label{appOnePoint} 
The one point function $f_N\left(L,\frac{L_1}{L}\right)$ can be computed in the same manner as the volumes $\mathcal{Z}_N$.
\begin{eqnarray}
f_N\left(L,\frac{L_1}{L}\right)
&=& \mathcal{Z}_{N+3}(L,L_1) \cr
&=& 2\sum_{d_1=0}^{+\infty}\left\langle\,\left(\frac{1}{2}\sum_{d}\frac{L^{2d}}{2^d d!}\tau_d\right)^{N+2}\frac{L_1^{2d_1}}{2^{d_1}d_1 !}\tau_{d_1} \,\right\rangle_0 \cr
&=& \frac{2}{2^{N+2}}\sum_{d_1,\dots,d_{N+3}}\frac{L^{2(d_2+\dots+d_{N+3})}}{2^{d_2+\dots+d_{N+3}}d_2 !\dots d_{N+3}!}\frac{L_1^{2d_1}}{2^{d_1}d_1!}\langle\tau_{d_1}\dots\tau_{d_{N+3}} \rangle_0 \cr 
\end{eqnarray}
Now:
\begin{equation}
\langle\tau_{d_1}\dots\tau_{d_{N+3}} \rangle_0=\frac{N!}{d_1!\dots d_{N+3}!}\delta(N-\sum_{i=1}^{N+3} d_i).
\end{equation}
So:
\begin{eqnarray}
f_N\left(L,\frac{L_1}{L}\right)
&=&\frac{N! L^{2N}}{2}\sum_{d_1+\dots d_{N+3}=N}\frac{1}{2^{2(d_2+\dots+d_{N+3})}}\left(\frac{L_1}{2L}\right)^{2d_1}\frac{1}{(d_1!\dots d_{N+3}!)^2} \cr 
&=&\frac{N!L^{2N}}{2}[z^{2N}]I_0(z)^{N+2}I_0\left(z\frac{L_1}{L}\right) \cr
&=&\frac{N!L^{2N}}{2}\underset{z\to 0}{\operatorname{Res}}\frac{dz}{z^{1+2N}}I_0(z)^{N+2}I_0\left(z\frac{L_1}{L}\right) \cr 
&=&\frac{N!L^{2N}}{2}\underset{z\to 0}{\operatorname{Res}}\frac{dz}{z}I_0(z)^2 e^{N(\ln{I_0(z)}-2\ln{z}+\frac{1}{N} I_0(zL_1/L))}.
\end{eqnarray}

\end{document}